%% file: feynRulesCTP.tex
\documentclass[prd, twocolumn, amsfonts, amsmath, amssymb, twoside, showpacs, nofootinbib,preprintnumbers]{revtex4}

\usepackage{bbm}
\usepackage{bm}
\usepackage{graphicx}
\usepackage{latexsym}
\usepackage{mathrsfs}

\input{feynRulesCTPInput}

\begin{document}

\title{\KB Equations with Non-Gaussian Initial Conditions:\\The Equilibrium Limit}

\author{Mathias Garny}
\email{Mathias.Garny@mpi-hd.mpg.de}
\affiliation{\MPIK}

\author{Markus Michael M\"uller}
\email{Markus.Michael.Mueller@lrz.de}
\affiliation{\LRZ}

\date{\today}
\preprint{Phys.~Rev.~{\bf D80} (2009) 085011}

\begin{abstract}
The nonequilibrium dynamics of quantum fields is an initial-value problem, which can be described
by \KB equations.
Typically, and in particular when numerical solutions are demanded, these \KB equations are restricted
to Gaussian initial states. However, physical initial states are non-Gaussian correlated initial
states. In particular, renormalizability requires the initial state to feature $n$-point correlations
that asymptotically agree with the vacuum correlations at short distances. In order to
identify physical nonequilibrium initial states, it is therefore a precondition to describe the
vacuum correlations of the interacting theory within the nonequilibrium framework.
In this paper, \KB equations for non-Gaussian correlated initial states describing vacuum and thermal
equilibrium are derived from the 2PI effective action. A diagrammatic method for the explicit
construction of vacuum and thermal initial correlations from the 2PI effective action is provided.
We present numerical solutions of \KB equations for a real scalar $\Phi^4$ quantum field theory which
take the thermal initial 4-point correlation as the leading non-Gaussian correction into account. We find that
this minimal non-Gaussian initial condition yields an approximation to the complete equilibrium initial
state that is quantitatively and qualitatively significantly improved as compared to Gaussian initial
states.
\end{abstract}

\maketitle

%
\section{Introduction}
%

	Nonequilibrium processes within astro-particle and high-energy physics, like reheating after
	inflation, baryogenesis, or relativistic heavy ion collisions, are typically described by
	classical or semi-classical equations. These include Boltzmann equations, hydrodynamic
	transport equations or effective equations of motion for a coherent scalar field expectation
	value~\cite{Kofman:1997yn,Kolb:2003dz,Kolb:1990vq}.
	The semi-classical treatment provides the possibility to relate fundamental parameters of
	the underlying theory with model predictions. Although inflation and baryogenesis occur at
	extremely high energies, key observables like the baryon asymmetry and the spectral index
	are subject to experimental verification, for example by measurements of the cosmic microwave
	background radiation~\cite{Komatsu:2008hk}.
	Therefore it is of great importance to assess the reliability of the underlying semi-classical
	approximations.
	This can be achieved by a comparison with a completely quantum field theoretical treatment.
	
	In recent years it has been demonstrated that the time evolution of relativistic scalar and
	fermionic quantum fields far from equilibrium can be described based on first principles by
	\KB equations~\cite{Berges:2000ur,Berges:2001fi,Aarts:2001yn,Aarts:2003bk,Lindner:2005kv}.
	These equations for the complete one- and two-point correlation functions can be obtained
	from the stationarity conditions of the 2PI effective action~\cite{Cornwall:1974vz} defined
	on the Schwinger-Keldysh closed real-time contour~\cite{Schwinger:1960qe,Bakshi:1962dv,%
	Bakshi:1963bn,Keldysh:1964ud,Danielewicz:1982kk}.
	The advantages of this approach are manifold:
	First, its conceptual simplicity is very attractive. The only assumption entering the
	derivation of \KB equations is the truncation of the so-called 2PI functional, which amounts
	to a controlled approximation in the coupling constant or the inverse number of field degrees
	of freedom for specific quantum field theories~\cite{Berges:2001fi}.
	Furthermore, \KB equations inherently
	incorporate typical quantum (e.g. off-shell) effects as well as `classical' (e.g. on-shell)
	effects in a unified manner, and can be applied even to systems far from equilibrium.
	Accordingly, they are very versatile and can be employed both to assess the validity of
	conventional semi-classical approximations (e.g. for baryogenesis and leptogenesis), and in
	situations where a single effective description does not exist (e.g. for (p)reheating by
	inflaton decay and subsequent thermalization)~\cite{Berges:2002cz,Arrizabalaga:2004iw,%
	Aarts:2007qu,Aarts:2007ye,Berges:2009bx}.
	
	It has been shown that numerical solutions of \KB equations not only provide a description of
	the quantum thermalization process of relativistic quantum fields for closed
	systems~\cite{Berges:2000ur,Berges:2002wr}, but also feature a separation of time-scales
	between kinetic and chemical equilibration (prethermalization)~\cite{Berges:2004ce}.
	Furthermore, they have been compared to semi-classical transport equations for bosonic and
	fermionic systems~\cite{Aarts:2001yn,Juchem:2003bi,Arrizabalaga:2005tf,Lindner:2005kv,%
	Lindner:2007am,Anisimov:2008dz} (see also Refs.~\cite{Danielewicz:1982ca,Kohler:1995zz,%
	Kohler:1996zz,Morawetz:1998em,Kohler:2001zv} for the non-relativistic case).
	Moreover, \KB equations can describe the decay of a coherent, oscillating scalar field
	expectation value under conditions where parametric resonance occurs~\cite{Berges:2002cz},
	and have also been investigated in curved space-time~\cite{Tranberg:2008ae,Hohenegger:2008zk}.
	
	These successes of the 2PI effective action and \KB equations in the area of nonequilibrium
	quantum field theory make it worthwhile and, in view of realistic applications, necessary to
	answer remaining conceptual questions, like renormalization.
	The renormalization of the 2PI effective action in vacuum and at finite temperature has been
	established
	recently~\cite{Berges:2005hc,Berges:2004hn,Blaizot:2003an,vanHees:2001pf,vanHees:2001ik}.
	It has been shown that the vacuum counterterms are also sufficient at finite temperature. In
	order to extend this proof to nonequilibrium situations, it is necessary to identify initial
	states that are themselves free of divergences. In particular, this requires that the
	correlation functions characterizing these \emph{physical} initial states are rendered finite
	by the vacuum counterterms.

	Typically, \KB equations are solved for Gaussian initial states. All connected $n$-point
	correlation functions with $n>2$ vanish for Gaussian initial states by definition. However,
	in vacuum and at finite temperature the $3$- and $4$-point correlation functions carry overall
	divergences that are cancelled by corresponding vacuum counterterms. Thus Gaussian initial
	states lead to an unbalanced divergence at the initial time~\cite{Borsanyi:2008ar}. In order
	to overcome this shortcoming, physical initial states have to carry non-Gaussian initial $3$-
	and $4$-point correlation functions that differ from the vacuum correlations at most by a
	finite amount.

	In order to be able to identify physical initial states, it is therefore a precondition to be
	able to describe vacuum and thermal equilibrium within the standard framework of
	nonequilibrium quantum field theory, i.e. on the closed real-time path with finite initial
	time $t_{init}\equiv 0$.
	Apart from the question of renormalization it is also a matter of principle that vacuum and
	thermal equilibrium should be accessible within nonequilibrium field theory as special cases
	by choosing the initial state appropriately.

	In this work, \KB equations for non-Gaus\-sian correlated initial states describing vacuum and
	thermal equilibrium are derived from the 2PI effective action formulated on the closed
	real-time path with finite initial time. For that purpose, we propose a diagrammatic method
	for the explicit construction of vacuum and thermal initial correlations that is
	applicable to nonperturbative 2PI approximations.
	
	There exist several techniques to describe non-Gaus\-sian correlated initial states.
	These can be divided into two categories: Either, the correlations are generated by modifying
	the closed real-time path $\C$, or the initial state is explicitly described by its density
	matrix~$\rho$. We shall refer to these as \emph{implicit} and \emph{explicit} techniques,
	respectively.
	The implicit techniques include the so-called
	imaginary-time stepping~\cite{Danielewicz:1982kk,Danielewicz:1982ca,Kohler:1995zz}, where an
	imaginary branch is added to the contour $\C$, similar to the description of thermal field
	theory~\cite{Niemi:1983ea,Niemi:1983nf,Landsman:1986uw}.
	Another possibility is to extend the closed real-time contour $\C$ over the complete real axis,
	such that it runs from $-\infty$ to $+\infty$, and back to $-\infty$. The correlated initial
	state is then generated by including an external two-point source $K(x,y)$, that is switched
	off at the `initial' time~\cite{Borsanyi:2008ar}.
	
	For the explicit technique, the density matrix $\rho$ of the initial state is parameterized by
	initial $n$-point correlation functions
	$\alpha_n(x_1,\dots,x_n)$~\cite{Chou:1984es,Calzetta:1986cq}. These appear in the form of
	non-local effective $n$-point vertices in the 2PI effective action. When deriving \KB
	equations, these inherit the contributions from non-Gaussian initial correlations.

	The advantage of the implicit techniques is that the equilibrium limit can be approached
	without any additional work. However, due to the implicit preparation of the initial state,
	the freedom and the control in choosing the initial state is restricted.
	The advantage of the explicit technique is that the resulting \KB equations are very similar
	to the Gaussian case. Furthermore, the explicit approach provides a maximal degree of freedom
	for specifying the initial state.
	
	The renormalization of \KB equations has recently been discussed based on the implicit
	technique involving an external two-point source~\cite{Borsanyi:2008ar}.
	In this work, we use the explicit technique. Thus, the methods developed in this paper
	provide a complementary framework for addressing the issue of renormalization.

	The main purpose of this work is to provide techniques for calculating the non-Gaussian initial
	correlations $\alpha_n^{\it th}(x_1,\dots,x_n)$ for a thermal initial state $\rho_{\it th}$
	within the 2PI-Schwinger-Keldysh formalism. As explained above, this is a prerequisite for
	studying the renormalization of \KB equations based on the explicit approach.
	In Ref.~\cite{Calzetta:1986cq}, a perturbative expansion of the $\alpha_n^{\it th}$ was
	derived. Unfortunately, this expansion is not suitable for the non-perturbative 2PI formalism.
	The main idea followed in the present work is to determine the $\alpha_n^{\it th}$ by matching
        the \KB equations for a thermal initial state on the one hand with the evolution equations
        obtained from the 2PI effective action formulated on the well-known thermal time
	contour~\cite{Niemi:1983ea,Niemi:1983nf,Landsman:1986uw,LeBellac:1996at,Gelis:1994dp} on the
	other hand.
	For this matching procedure, we employ the thermal time contour obtained by concatenating the
	closed real-time contour with the imaginary time contour. We stress that it is important to
	keep a finite `initial' time. In this case, both the horizontal and the vertical branches
	contribute~\cite{LeBellac:1996at,Gelis:1994dp,Gelis:1999nx}. The contributions from the vertical
	branch can then be identified with corresponding contributions from the $\alpha_n^{\it th}$
	within the equivalent Schwinger-Keldysh formalism. We note that, when considering an exact
	thermal initial state, the propagator obtained from the \KB equations is time-translation
	invariant within its domain of definition, in accordance with
	Refs.~\cite{LeBellac:1996at,Gelis:1994dp,Gelis:1999nx}.
	\begin{figure}
		\includegraphics[width=0.6\columnwidth,keepaspectratio]{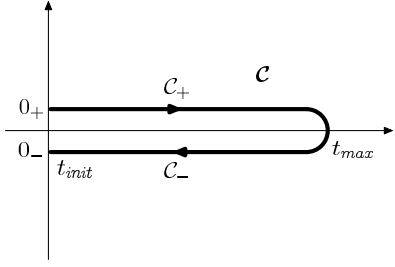}
		\caption{\label{ROOE:fig:CTP}\captionstyle%
			Closed real-time path $\C$. This time path was invented by
			Schwinger~\cite{Schwinger:1960qe} and applied to non-equilibrium problems by
			Keldysh~\cite{Keldysh:1964ud} (see also Refs.~\cite{Bakshi:1962dv,%
			Bakshi:1963bn}). In order to avoid the doubling of the degrees of freedom, we
			use the form presented in Ref.~\cite{Danielewicz:1982kk}.
		}
	\end{figure}

	This work is organized as follows:
	In section~\ref{ROOE:sec:EAaKBE}, we derive \KB equations for non-Gaussian initial states from
	the corresponding 2PI effective action using the explicit technique.
	In section~\ref{ROOE:sec:PerturbativeThermInitCorr}, we provide techniques for calculating
	the non-Gaussian initial correlations $\alpha_n^{\it th}(x_1,\dots,x_n)$ within perturbation theory.
	These techniques are generalized to the nonperturbative 2PI case in
	section~\ref{ROOE:sec:NonPerturbativeThermInitCorr}.
	In section~\ref{ROOE:sec:KBEThermalCTP}, \KB equations for a thermal initial state are
	derived by combining the results from sections~\ref{ROOE:sec:EAaKBE}
	and~\ref{ROOE:sec:NonPerturbativeThermInitCorr}.
	In section~\ref{ROOE:sec:Numerics}, we compare numerical solutions of \KB equations for two
	nonequilibrium initial states that are approximations to the thermal initial state. One of them
	is Gaussian, while the other also includes the leading non-Gaussian initial correlation.
	The appendices~\ref{QFOOE:sec:ThermalInitialState} and~\ref{QFOOE:sec:FullConnection} contain
	additional material helpful for sections~\ref{ROOE:sec:PerturbativeThermInitCorr}
	and~\ref{ROOE:sec:NonPerturbativeThermInitCorr}, respectively.

%
\section{Effective Action and \KB equations}\label{ROOE:sec:EAaKBE}
%
	
	\subsection{Gaussian Initial States}

	In this subsection, we start from the classical action for a real scalar quantum field with a
	quartic self interaction
	\begin{equation}\label{ROOE:Action}
		S[\phi] = \int d^4x \, \left( \frac{1}{2} ( \partial\phi )^2 - \frac{1}{2} m^2 \phi^2
			- \frac{\lambda}{4!} \phi^4 \right) \;,
	\end{equation}
	and review the basic elements of the derivation of the 2PI effective action and the \KB
	equations for the case of a Gaussian initial state. In the following subsections, we can
	then easily expose the differences, which arise for a non-Gaussian initial state.

	The Schwinger-Keldysh propagator is defined by
	\begin{equation} \label{ROOE:SchwingerKeldyshPropagator}
		G(x,y) = \left\langle T_\C\, \Phi(x) \Phi(y) \right\rangle
		- \left\langle \Phi(x) \right\rangle \left\langle \Phi(y) \right\rangle \;,
	\end{equation}
	where $T_\C$ denotes the time-ordering operator along the closed real-time path $\C$ shown
	in figure~\ref{ROOE:fig:CTP} \cite{Schwinger:1960qe,Keldysh:1964ud,Danielewicz:1982kk}.
	The Schwinger-Keldysh propagator can be obtained by functional differentiation from the
	generating functional for correlation functions formulated on the closed real-time path.
	The generating functional in the presence of a local external source $J(x)$ and a bilocal
	external source $K(x,y)$ is given by \cite{Calzetta:1986cq}
	\begin{eqnarray}
		\lefteqn{Z_{\rho} \left[ J, K \right] = \int \mathcal{D} \varphi \left( x \right) \left\langle \varphi_+ \left| \rho \right| \varphi_- \right\rangle} \nonumber \\
		& & {} \times \exp \left( i S \left[ \varphi \right] + i J \varphi + \frac{i}{2} \varphi K \varphi \right) \;, \label{ROOE:GeneratingFunctional}
	\end{eqnarray}
	where a matrix-vector notation has been used for the space-time integrals in the exponential
	function%
	\footnote{%
		Throughout this work, the compact notation of Ref.~\cite{Danielewicz:1982kk} is used
		for the contour integrals over the closed real-time path, for example
		$
			J\varphi \equiv \int_\C\! d^4x \, J(x)\varphi(x)
			= \int\! d^4x \left[ J_+(x)\varphi_+(x) - J_-(x)\varphi_-(x) \right]
		$.
	}%
	, and $|\varphi_\pm \rangle$ are the quantum states corresponding to the
	field configurations $\varphi_\pm(\bm{x}) = \varphi(0_\pm,\bm{x})$. The information about the
	initial state enters via the matrix element of the density matrix $\rho$, which is known only
	at the initial time $t = t_{init} \equiv 0$.

	A Gaussian initial state is an initial state for which all connected $n$-point correlation
	functions with $n \geq 3$ vanish at the initial time. The density matrix element for a
	Gaussian initial state can be parameterized by
	\begin{equation}\label{ROOE:GaussianDensityMatrixElement}
		\left\langle \varphi_+ \left| \rho \right| \varphi_- \right\rangle
		= \exp\left(i \alpha_0 + i \alpha_1 \varphi + \frac{i}{2} \varphi \alpha_2 \varphi \right) \;.
	\end{equation}
	Therefore, in the Gaussian case, the contribution of the density matrix to the generating
	functional~(\ref{ROOE:GeneratingFunctional}) can formally be absorbed into the external sources,
	$J+\alpha_1\rightarrow J$ and $K+\alpha_2\rightarrow K$.
	As is, for example, explained in Ref.~\cite{Berges:2004yj}, this means that the Gaussian
        initial density matrix does not appear explicitly, but rather enters the dynamics via defining
        the initial conditions of all independent one- and two-point functions. For vanishing field
        expectation value, these are given by $G(x,y)$, $(\partial_{x^0}+\partial_{y^0})G(x,y)$ and
	$\partial_{x^0}\partial_{y^0}G(x,y)$, all evaluated at
	$x^0=y^0=0$~\cite{Berges:2004yj,Lindner:2005kv}.

	The 2PI effective action $\Gamma[\phi,G]$ is the double Legendre transform of the generating
	functional~(\ref{ROOE:GeneratingFunctional}) with respect to the external sources. For a
	Gaussian initial state, the generating functional has the same structure as the generating
	functional in vacuum, except that all time-integrations are performed over the closed
	real-time path.
	Consequently, for a Gaussian initial state the 2PI effective action can be parameterized
	in the form~\cite{Cornwall:1974vz}
	\begin{eqnarray}
		\Gamma \left[ \phi, G \right] & = & S \left[ \phi \right] + \frac{i}{2} \Tr \log_{\C} \left[ G^{-1} \right] \nonumber \\
		                              &   & {} + \frac{i}{2} \Tr_{\C} \left[ \mathcal{G}_0^{-1} G \right] + \Gamma_2 \left[ \phi, G \right] \;, \label{ROOE:2PIEffAct}
	\end{eqnarray}
	where ${\mathcal{G}_0}^{-1}$ is the inverse classical Schwinger-Keldysh propagator and 
	$i\Gamma_2[\phi,G]$ is the sum of all 2PI Feynman diagrams without any external legs, where
	internal lines represent the complete Schwinger-Keldysh propagator $G(x,y)$. The vertices of
	the diagrams contained in $i\Gamma_2[\phi,G]$ are given by the third and fourth functional
	derivatives of the classical action $S[\phi]$~\cite{Cornwall:1974vz}. Eventually, the \KB
	equations
	\begin{eqnarray}
		\left( \Box_x + M^2(x) \right) G_F (x,y)      & = & \intl_{0}^{y^0} d^4 z \; \Pi_F(x,z) G_\rho(z,y) \nonumber \\
							      &   & \hspace{-15mm} {} - \intl_{0}^{x^0} d^4 z \; \Pi_\rho(x,z) G_F(z,y) \;, \label{ROOE:KBEGaussGFGRho} \\
		\left( \Box_x + M^2(x) \right) G_{\rho} (x,y) & = & \intl_{x^0}^{y^0} d^4 z \; \Pi_\rho(x,z) G_\rho(z,y) \;, \nonumber
	\end{eqnarray}
	follow from the stationarity condition of the 2PI effective action. Here, we use the notation
	of Ref.~\cite{Lindner:2005kv}.
	
	\subsection{Non-Gaussian Initial States}
	
	In the remainder of this section, \KB equations are derived that can describe systems
	characterized by a general non-Gaussian initial
	state~\cite{Calzetta:1986cq,Chou:1984es,Garny:PhD}. For that purpose, we extend the derivation
	of the previous subsection using a generalization of the Gaussian density
	matrix~(\ref{ROOE:GaussianDensityMatrixElement}). In general, the matrix element of the
	density matrix $\rho$ is an arbitrary functional of the field configurations
	$\varphi_+(\bm{x})$ and $\varphi_-(\bm{x})$, which can be written as~\cite{Calzetta:1986cq}
	\begin{equation}\label{ROOE:DensityMatrixElement}
		\left\langle \varphi_+ \left| \rho \right| \varphi_- \right\rangle
		= \exp\left( i F[\varphi] \right) \;.
	\end{equation}
	While for a Gaussian initial state $F[\varphi]$ is a quadratic functional of the field, for a
	general non-Gaussian initial state it may be Taylor expanded in the form~\cite{Calzetta:1986cq}
	\begin{eqnarray}
	  F[\varphi] & = & \sum_{n=0}^\infty \frac{1}{n!} \intl_{\C} d^4 x_1 \ldots d^4 x_n \; \alpha_n \left( x_1, \ldots, x_n \right) \nonumber \\
	             &   & {} \times \varphi(x_1) \cdot \ldots \cdot \varphi(x_n) \;. \label{ROOE:ExpansionOfDensityMatrixElement}
	\end{eqnarray}
	By definition $F[\varphi]$ depends only on the field configuration evaluated at the boundaries
	of the time contour. Consequently, the kernels $\alpha_n(x_1,\dots,x_n)$ are non-zero only if
	all their time arguments lie on the boundaries of the time contour. With the notation
	$\delta_+(t)=\delta_\C(t-0_+)$ and $\delta_-(t)=\delta_\C(t-0_-)$, they can be written in the
	form
	\begin{eqnarray}
	        \alpha_n \left( x_1, \ldots, x_n \right)
	  & = & \alpha_n^{\epsilon_1, \ldots, \epsilon_n} \left( \bm{x_1}, \ldots, \bm{x_n} \right) \nonumber \\
	  &   & {} \times \delta_{\epsilon_1} \left( x_1^0 \right) \cdot \ldots \cdot \delta_{\epsilon_n} \left( x_n^0 \right) \;, \label{ROOE:AlphaSource}
	\end{eqnarray}
	where summation over $\epsilon_j \in \{+,-\}$ is implied. In this way, the explicit dependence
	of the functional $F[\varphi]$ on the field configurations $\varphi_+(\bm{x})$ and
	$\varphi_-(\bm{x})$ may be recovered,
	\[ F[\varphi] = \alpha_0 + \int d^3 x \; \alpha_1^\epsilon (\bm{x}) \varphi_{\epsilon} (\bm{x}) + \ldots \;. \]
	The set of all kernels $\alpha_n$ with $n\geq 0$ encodes the complete information about the
	density matrix characterizing the initial state. Not all the kernels are independent. The
	Hermiticity of the density matrix, $\rho=\rho^\dag$, implies that
	\[
		i\alpha_n^{\epsilon_1,\dots,\epsilon_n}(\bm{x_1},\dots,\bm{x_n}) = \left( i\alpha_n^{(-\epsilon_1),\dots,(-\epsilon_n)}(\bm{x_1},\dots,\bm{x_n}) \right)^* \;.
	\]
	If the initial state is invariant under some symmetries, there are further constraints.
	For example, for an initial state which is invariant under the $Z_2$-symmetry
	$\Phi\rightarrow-\Phi$, all kernels $\alpha_n(x_1,\dots,x_n)$ with odd $n$ vanish. If the
	initial state is homogeneous in space, the initial correlations $\alpha_n(x_1,\dots,x_n)$ are
	invariant under space-translations $\bm{x_i}\rightarrow\bm{x_i}+\bm{a}$ of all arguments for
	any real three-vector $\bm{a}$, and can conveniently be expressed in spatial momentum space,
	\begin{eqnarray*}
		      \lefteqn{\alpha_n^{\epsilon_1,\ldots,\epsilon_n}(\bm{x_1},\ldots,\bm{x_n})} \\
		& = & \int \frac{d^3 k_1}{(2\pi)^{3}} \ldots \int \frac{d^3 k_n}{(2\pi)^{3}} \; \exp \left( i \sum_j \bm{k_j x_j} \right) \\
		&   & {} \times (2\pi)^3 \delta^3(\bm{k_1}+\dots+\bm{k_n}) \alpha_n^{\epsilon_1,\dots,\epsilon_n} (\bm{k_1},\dots,\bm{k_n}) \;.
	\end{eqnarray*}
	Summarizing, the generating functional reads
	\begin{eqnarray}
		Z_\rho[J,K] & = & \int \mathcal{D} \varphi \; \exp \Big( iS[\varphi] \nonumber \\
		            &   & {} + iJ\varphi + \frac{i}{2}\,\varphi K\varphi + iF_3[\varphi] \Big) \label{ROOE:BilocGeneratingFunctionalNonequilibrium} \;,
	\end{eqnarray}
	where the kernels $\alpha_0$, $\alpha_1$ and $\alpha_2$ have been absorbed into the measure
	$\mathcal{D}{\varphi}$ and into the sources $J$ and $K$, respectively. The functional
	$F_3[\varphi]\equiv F_3[\varphi;\alpha_3,\alpha_4,\dots]$ contains the contributions of
	third, fourth and higher orders of the Taylor
	expansion~(\ref{ROOE:ExpansionOfDensityMatrixElement}) and vanishes for a Gaussian initial state.
	
	\subsection[2PI Effective Action]{2PI Effective Action for Non-Gaussian Initial States}
	
\begin{figure}
		\feyn{0mm}{1.0}{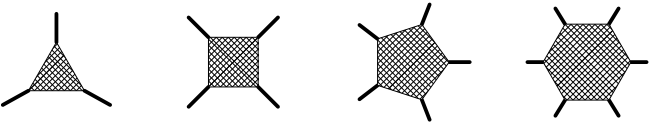}
		\caption{\label{ROOE:fig:VerticesFromInitialSource}\captionstyle%
			Non-local effective vertices $i\bar\alpha_n(x_1,\dots,x_n)$
			connecting $n$ lines for $n = 3, 4, 5, 6$
			encoding the non-Gaussian three-, four-, five-, and six-point correlations
			of the initial state.
		}
	\end{figure}

	According to eq.~(\ref{ROOE:BilocGeneratingFunctionalNonequilibrium}), the 2PI effective
	action in the presence of non-Gaussian correlations is obtained from the standard
	parameterization~\cite{Cornwall:1974vz} of the 2PI effective action applied to a theory
	described by the modified classical action $ \tilde S[\phi] \equiv S[\phi] + F_3[\phi],$
	\begin{eqnarray}\label{ROOE:ParamOf2PIEffActTilde}
		\Gamma \left[ \phi, G \right] &   =    & \tilde S \left[ \phi \right] + \frac{i}{2} \Tr \log_{\C} \left[ G^{-1} \right] \nonumber \\
			                      &        & {} + \frac{i}{2} \Tr_{\C} \left[ \tilde{ \mathcal{G} }_0^{-1} G \right] + \tilde \Gamma_2 \left[ \phi, G \right] \nonumber \\
			                      & \equiv & \Gamma_G[\phi,G] + \Gamma_{nG}[\phi,G] \;,
	\end{eqnarray}
	where
	$i\tilde{\mathcal{G}}_0^{-1}(x,y) \equiv \delta^2\tilde S[\phi]/\delta\phi(x)\delta\phi(y)$.
	The Gaussian part $\Gamma_G[\phi,G]$ coincides with the right-hand side of
	eq.~(\ref{ROOE:2PIEffAct}), and the non-Gaussian part is given by
	\[
		\Gamma_{nG}[\phi,G] = F_3[\phi]
		+ \frac{1}{2} \Tr_\C \left[ \frac{\delta^2F_3}{\delta\phi\delta\phi}G \right]
		+ \Gamma_{2,\,nG}[\phi,G]\,,
	\]
	where
	\[ \Gamma_{2,\,nG}[\phi,G]\equiv \tilde\Gamma_2[\phi,G] - \Gamma_{2}[\phi,G] \;. \]
	The modified 2PI functional $i\tilde\Gamma_2[\phi,G]$ is equal to the sum of all 2PI Feynman
	diagrams without any external legs, where internal lines represent the complete propagator
	$G(x,y)$ and where vertices are given by the functional derivatives of the modified classical
	action $\tilde S[\phi] = S[\phi] + F_3[\phi]$. The contribution from the classical action
	$S[\phi]$ leads to the classical local three- and four-point vertices. Additionally, the
	contribution from the functional $F_3[\phi]$ leads to effective non-local vertices, which
	contain the non-Gaussian initial $n$-point correlations with $n\geq 3$ (see
	figure~\ref{ROOE:fig:VerticesFromInitialSource}),
	\begin{equation}\label{ROOE:NonGaussianVertices}
		i\frac{\delta^n F_3[\phi]}{\delta\phi(x_1)\dots\delta\phi(x_n)}  \equiv
		i\bar\alpha_n(x_1,\dots,x_n).
	\end{equation}
	These effective $n$-point vertices are only supported at the initial time, and can be
	parameterized analogously to eq.~(\ref{ROOE:AlphaSource}). For a $Z_2$-symmetric initial
	state, the field expectation value vanishes, $\phi(x)=0$, such that
	$\bar\alpha_n(x_1,\dots,x_n)=\alpha_n(x_1,\dots,x_n)$.
	The contribution of these effective non-local vertices is most important close to the initial
	time. For example, a non-zero four-point source $\alpha_4(x_1,\dots,x_4)$ leads to a
	non-vanishing value of the connected four-point correlation function at the initial time,
	which is impossible for a Gaussian initial state.
	
	Note that those 2PI diagrams that contain exclusively the classical vertices contribute to the
        functional $i\Gamma_2[\phi,G]$ by definition. Therefore, the diagrams contributing to the
	non-Gaussian part $i\Gamma_{2,\,nG}[\phi,G]$ contain at least one effective vertex from
	eq.~(\ref{ROOE:NonGaussianVertices}).

\begin{figure}
		\feyn{0mm}{1.0}{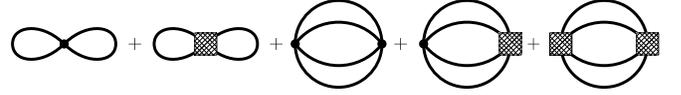}
		\caption{\label{ROOE:fig:Gamma2ThreeLoopWithInitial4PointSource}\captionstyle
			Diagrams contributing to the three-loop truncation
			of the 2PI effective action in the symmetric phase
			(setting-sun approximation) in the presence of an
			effective non-local four-point vertex.
		}
	\end{figure}

	In section \ref{ROOE:sec:Numerics} we study the numerical solution of \KB equations for a
	$Z_2$-symmetric non-Gaussian initial state with a non-zero initial 4-point correlation. In
	this case the 2PI functional $\tilde \Gamma_2$ reads in ``na\"{\i}ve''\footnote{This means
	that non-local effective vertices do not affect	the counting of loops.}	three-loop
	approximation (see figure~\ref{ROOE:fig:Gamma2ThreeLoopWithInitial4PointSource})
	\begin{eqnarray}
	  \lefteqn{i \tilde{\Gamma}_2[G] = \frac{1}{8} \intl_{\C} d^4 x_{1234} \left[ \rule{0mm}{4mm} -i\lambda\delta_{12}\delta_{23}\delta_{34} + i\alpha_{1234} \right]
		      G_{12}G_{34}} \nonumber \\
		&   & {} + \frac{1}{48} \intl_{\C} d^4 x_{1\cdots 8} \left[ \rule{0mm}{4mm} -i\lambda\delta_{12}\delta_{23}\delta_{34} + i\alpha_{1234} \right] 
		      \label{ROOE:Gamma2ThreeLoopWithInitial4PointSource} \\
		&   & {} \times \, G_{15}G_{26}G_{37}G_{48} \left[ \rule{0mm}{4mm} - i\lambda\delta_{56}\delta_{57}\delta_{58} + i\alpha_{5678} \right] \;, \nonumber \hspace{10mm}
	\end{eqnarray}
	where $G_{12}=G(x_1,x_2)$ and $\alpha_{1234}=\alpha_4(x_1,x_2,x_3,x_4)$. Note that the
	contribution to the mixed \emph{basketball} diagram in the second and third line with one
	classical and one effective vertex appears twice, which accounts for the symmetry factor
	$1/24$.
	%

	\subsection[Self-Energy]{Self-Energy for Non-Gaussian Initial States}

	The equation of motion for the complete propagator obtained from
	eq.~(\ref{ROOE:ParamOf2PIEffActTilde}) reads
	\begin{equation}\label{ROOE:Schwinger-Dyson}
	 	G^{-1}(x,y) = \mathcal{G}_0^{-1}(x,y) - \Pi(x,y) - i\bar\alpha_2(x,y)\,,
	\end{equation}
	where $\bar\alpha_2=\alpha_2+\delta^2F_3[\phi]/\delta\phi\delta\phi$ and the complete
	self-energy is given by
	\begin{eqnarray}
		\Pi(x,y) & = & \frac{2i\delta\tilde\Gamma_2[\phi,G]}{\delta G(y,x)} \nonumber \\
			 & = & \frac{2i\delta\Gamma_2[\phi,G]}{\delta G(y,x)} + \frac{2i\delta\Gamma_{2,\,nG}[\phi,G]}{\delta G(y,x)} \nonumber \\
			 &\equiv& \Pi^G(x,y) + \Pi^{nG}(x,y) \, , \label{ROOE:DefOfSelfEnergyNonGauss}
	\end{eqnarray}
	where $\Pi^G$ contains the contributions to the self energy, which are also present
	for a Gaussian initial state, and the non-Gaussian part $\Pi^{nG}$ contains diagrams
	with at least one non-local effective vertex. They can be further decomposed as
	\begin{eqnarray}\label{ROOE:SelfEnergyDecomposition}
		\Pi^{\it G}(x,y)  & = & -i\Pi_{\it loc}(x)\delta_\C(x-y) + \Pi^{\it G}_{\it non-loc}(x,y)\,, \nonumber \\
		\Pi^{\it nG}(x,y) & = & i\Pi^{\it nG}_{\it surface}(x,y) + \Pi^{\it nG}_{\it non-loc}(x,y)\,.
	\end{eqnarray}
	The non-Gaussian non-local part $\Pi^{\it nG}_{\it non-loc}(x,y)$ contains diagrams where
	both external lines are attached to a standard vertex. The non-Gaussian ``surface'' part
	$i\Pi^{\it nG}_{\it surface}(x,y)$ contains diagrams where at least one external line is
	attached to a non-local effective vertex. Thus, the surface part is supported only at the
	initial time surface where $x^0=0$ or $y^0=0$. In general, such contributions can
	arise in the following ways:
	\begin{enumerate}
		\item\label{ROOE:2} From diagrams where both external lines are connected to an
			effective non-local vertex as given in eq.~(\ref{ROOE:NonGaussianVertices}).
			They are supported at $x^0=y^0=0$.
		\item\label{ROOE:3} From diagrams where one of the two external lines is connected
			to an effective non-local vertex, while the other one is connected to a
			classical local vertex.
			They are supported at $x^0=0,y^0\geq 0$ or vice-versa.
		\item\label{ROOE:1} Via the contribution $i\bar\alpha_2(x,y)$ of the initial
			two-point source which is supported at $x^0=y^0=0$.
			This is the {\it only} Gaussian surface-contribution.
	\end{enumerate}
	\begin{figure}
		\feyn{0mm}{1.0}{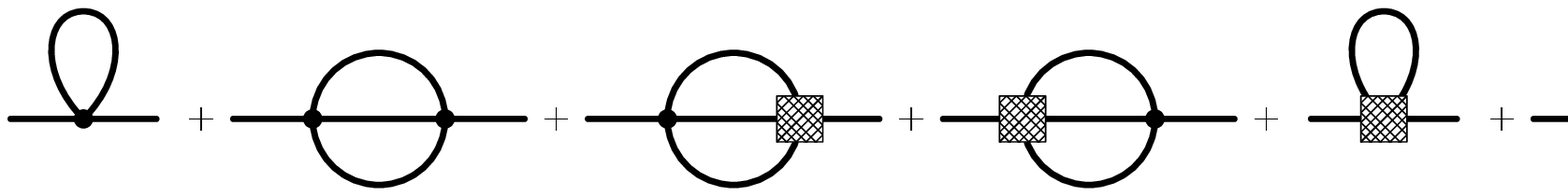}
		\caption{\label{ROOE:fig:PiThreeLoopWithInitial4PointSource}\captionstyle%
			Diagrams contributing to the self-energy $\Pi(x,y)$ in setting-sun
			approximation in the presence of an effective non-local four-point vertex.
			From left to right, the diagrams contribute to $\Pi_{\it loc}$,
			$\Pi^G_{\it non-loc}$, $\Pi_{\lambda\alpha}$, $\Pi_{\alpha\lambda}$, and the
			last two both contribute to $\Pi_{\alpha\alpha}$.
		}
	\end{figure}%
	Accordingly, the contributions to the self-energy which are supported at the initial time
	surface can be further decomposed as
	\begin{eqnarray*}
		\Pi_{\it surface}(x,y) & =      & \Pi^{\it nG}_{\it surface}(x,y) + \bar\alpha_2(x,y) \\
		                       & \equiv & \Pi_{\alpha\alpha}(x,y) + \Pi_{\lambda\alpha}(x,y) + \Pi_{\alpha\lambda}(x,y), \nonumber
	\end{eqnarray*}
	where
	\begin{eqnarray}\label{ROOE:PiSurfaceParametrization}
		\Pi_{\alpha\alpha}(x,y)  & = & \delta_{\epsilon_1}(x^0)\Pi_{\alpha\alpha}^{\epsilon_1,\epsilon_2}(\bm{x},\bm{y})\delta_{\epsilon_2}(y^0) \,, \nonumber\\
		\Pi_{\lambda\alpha}(x,y) & = & \Pi_{\lambda\alpha}^{\epsilon}(x^0,\bm{x},\bm{y})\delta_{\epsilon}(y^0) \,, \\
		\Pi_{\alpha\lambda}(x,y) & = & \delta_{\epsilon}(x^0)\Pi_{\alpha\lambda}^{\epsilon}(\bm{x},y^0,\bm{y}) \ = \ \Pi_{\lambda\alpha}(y,x) \,. \nonumber
	\end{eqnarray}
	$\Pi_{\alpha\alpha}$ contains all contributions of type (\ref{ROOE:2}.) and (\ref{ROOE:1}.).
	Diagrams of type (\ref{ROOE:3}.) contribute to $\Pi_{\lambda\alpha}$ or $\Pi_{\alpha\lambda}$
	depending which external line is attached to the effective non-local vertex and which to the
	classical local vertex. For all diagrams contributing to $\Pi_{\lambda\alpha}$ the left line
	is connected to the classical four- or three-point vertex.
	The non-local part of the self-energy can be split into statistical and spectral components,
	similarly to the Gaussian case,
	\begin{eqnarray}\label{ROOE:PiNonLocNonGauss}
		\Pi_{\it non-loc}(x,y)
			& \equiv & \Pi^{\it G}_{\it non-loc}(x,y) + \Pi^{\it nG}_{\it non-loc}(x,y) \\
			& \equiv & \Pi_F(x,y) - \frac{i}{2}\,\sgn_\C(x^0-y^0)\,\Pi_\rho(x,y) \,. \nonumber
	\end{eqnarray}
	The local part of the self-energy is identical to the Gaussian case and is included in the
	effective time-dependent mass term $M^2(x) = m^2 + \Pi_{loc}(x)$.
	
	For the setting-sun approximation from eq.~(\ref{ROOE:Gamma2ThreeLoopWithInitial4PointSource}),
	the self-energy is given by (see figure~\ref{ROOE:fig:PiThreeLoopWithInitial4PointSource})
	\begin{eqnarray*}
		\Pi^{G}_{non-loc}(x,y)  & = & \frac{(-i\lambda)^2}{6}G(x,y)^3 \, , \\
		\Pi^{nG}_{non-loc}(x,y) & = & 0 \, ,
	\end{eqnarray*}
	\begin{eqnarray*}
		i\Pi_{\alpha\alpha}(x,y)  & = & i\alpha_2(x,y) + \frac{1}{2}\int\!\!d^4\!x_{34}\,i\alpha_{xy34}G_{34} \nonumber \\
		\lefteqn{
			{} + \frac{1}{6}\int\!\!d^4\!x_{2\cdots7}\, i\alpha_{x234}G_{25}G_{36}G_{47}\,i\alpha_{567y} \, ,
		} \nonumber \\
		i\Pi_{\lambda\alpha}(x,y) & = & \frac{-i\lambda}{6}\int\!\!d^4\!x_{123} \, G_{x1}G_{x2}G_{x3}\,i\alpha_{123y} \, ,
	\end{eqnarray*}
	\begin{eqnarray*}
		i\Pi_{\alpha\lambda}(x,y) & = & \frac{-i\lambda}{6}\int\!\!d^4\!x_{234} \, i\alpha_{x234}\,G_{2y}G_{3y}G_{4y} \, .
	\end{eqnarray*}

	\subsection{\KB Equations for Non-Gaussian Initial States}
	
	Convoluting eq.~(\ref{ROOE:Schwinger-Dyson}) with the complete propagator yields
	\begin{eqnarray}
		\lefteqn{\left( \Box_x + M^2(x) \right) G(x,y) = -i \delta_{\C}(x-y)} \label{ROOE:KBENonGauss} \\
		& & {} -i \intl_{\C} d^4 z \; \left[ \Pi_{non-loc}(x,z) + i\Pi_{\lambda\alpha}(x,z) \right] G(z,y) \;. \nonumber
	\end{eqnarray}
	The second line follows from the parameterization~(\ref{ROOE:SelfEnergyDecomposition})
	of the self-energy, and assuming $x^0>0$ and $y^0>0$. Using
	eqs.~(\ref{ROOE:PiSurfaceParametrization},\ref{ROOE:PiNonLocNonGauss}) and transforming to
	spatial momentum space yields the \KB equations for $G_F(x^0,y^0,\bm{k})$ and
	$G_\rho(x^0,y^0,\bm{k})$  for spatially homogeneous non-Gaussian initial states,
	\begin{widetext}
	\begin{eqnarray}
	        \left( \partial_{x^0}^2 + \bm{k}^2 + M^2(x^0) \right) G_F (x^0, y^0, \bm{k})
	  & = & \intl_{0}^{y^0}\!\! dz^0\, \Pi_F(x^0, z^0, \bm{k}) G_\rho(z^0, y^0, \bm{k}) - \intl_{0}^{x^0}\!\!dz^0\, \Pi_\rho(x^0, z^0, \bm{k}) G_F(z^0, y^0, \bm{k}) \nonumber \\
	  &   & {} +\Pi_{\lambda\alpha,F} (x^0, \bm{k}) G_F (0, y^0, \bm{k}) + \frac{1}{4} \Pi_{\lambda\alpha,\rho} (x^0, \bm{k}) G_\rho (0, y^0, \bm{k}) \;, \label{ROOE:KBENonGaussGFGRho}
	\end{eqnarray}
	and
	\begin{eqnarray}
	        \left( \partial_{x^0}^2+ \bm{k}^2 + M^2(x^0) \right) G_\rho(x^0,y^0,\bm{k})
	  & = & \intl_{x_0}^{y^0}\!\! dz^0\, \Pi_\rho(x^0,z^0,\bm{k})  G_\rho(z^0,y^0,\bm{k})\,, \nonumber
	\end{eqnarray}
	\end{widetext}
	where
	\begin{eqnarray*}
		\Pi_{\lambda\alpha,F}(x^0,\bm{k})    & = &  \Pi_{\lambda\alpha}^+(x^0,\bm{k}) + \Pi_{\lambda\alpha}^-(x^0,\bm{k}) \,, \nonumber\\
		\Pi_{\lambda\alpha,\rho}(x^0,\bm{k}) & = &  2i\left( \Pi_{\lambda\alpha}^+(x^0,\bm{k}) - \Pi_{\lambda\alpha}^-(x^0,\bm{k}) \right) \,.
	\end{eqnarray*}
	Thus, for a non-Gaussian initial state, the right-hand side of the \KB equation for the
	statistical propagator is modified. In addition to the memory integrals there are now 
	new contributions originating from the non-Gaussian initial correlations.
	Unlike the memory integrals, these new contributions do \emph{not} have to vanish in the limit
	$x^0,y^0\rightarrow 0$. This is due to the fact that the higher non-Gaussian correlations of
	the initial state can lead to a non-vanishing value of the connected four- and three-point
	correlation functions at the initial time.

%
\section{Thermal Initial Correlations: Perturbation Theory}\label{ROOE:sec:PerturbativeThermInitCorr}
%

	%
	\begin{figure}
	\centering
		\includegraphics[width=0.6\columnwidth,keepaspectratio]{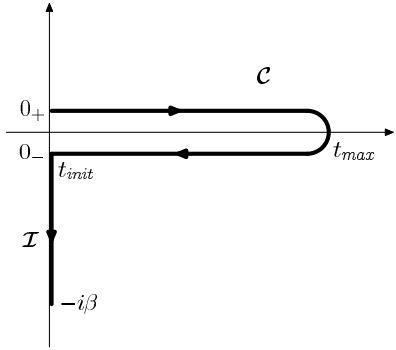}
		\caption{\label{ROOE:fig:TTP}\captionstyle%
			The thermal time contour $\CpI$ is obtained by concatenating the closed
			real-time contour $\C$ and the imaginary time contour $\I$ running from $t=0$
			to $t=-i\beta$~\cite{Niemi:1983ea,Niemi:1983nf,Landsman:1986uw,%
			LeBellac:1996at,Gelis:1994dp}. We employ this time path in order to
			infer the initial correlation functions $\alpha_n^{\it th}(x_1,\dots,x_n)$
			required for describing thermal equilibrium on the Schwinger-Keldysh
			closed real-time path $\C$ with finite initial time shown in
			figure~\ref{ROOE:fig:CTP}.
		}
	\end{figure}
	
	In order to derive \KB equations that are capable of describing thermal equilibrium, the
	thermal density matrix
	\[
		\rho_{th} = \frac{1}{Z} \, \exp\left(-\beta H\right) \,,
	\]
	has to be represented by a Taylor expansion in terms of thermal correlation functions
	$\alpha_n^{\it th}(x_1,\dots,x_n)$ as in
	eqs.~(\ref{ROOE:DensityMatrixElement},\,\ref{ROOE:ExpansionOfDensityMatrixElement}). These
	thermal correlation functions do then enter the \KB equations in the form of non-local
	effective vertices, as described in the previous section.

	The thermal correlations functions $\alpha_n^{\it th}(x_1,\dots,x_n)$ can be calculated
	order-by-order in the coupling constant within usual perturbation theory (see
	appendix~\ref{QFOOE:sec:ThermalInitialState}). However, in the context of \KB equations,
	it is necessary	to use approximations of the thermal correlation functions that are
	compatible with the underlying truncation of the 2PI effective action.
	It is a major purpose of this paper to provide computational techniques for identifying
	suitable approximations of the thermal correlation functions
	$\alpha_n^{\it th}(x_1,\dots,x_n)$.
	
	For simplicity, in this section, we first present the computational techniques within perturbation
	theory. In the following section, these techniques are then generalized to the nonperturbative
	2PI formalism.

	The main idea is to determine the functions $\alpha_n^{\it th}(x_1,\dots,x_n)$
	by matching the description of thermal equilibrium based on the closed real-time
	path $\C$ in the presence of effective vertices $\alpha_n^{\it th}$
	(``\,$\CpA$\,'') on the one hand with the well-known equivalent description based on the
	thermal time path (``\,$\CpI$\,'')~\cite{Niemi:1983ea,Niemi:1983nf,Landsman:1986uw,%
	LeBellac:1996at,Gelis:1994dp} shown in figure~\ref{ROOE:fig:TTP} on the other hand.

	The generating functional for these two descriptions is obtained by inserting the respective
	representations of the thermal density matrix,
	\begin{equation}\label{QFOOE:ThermalDensityMatrixElement}
		\left\langle \varphi_+ \left| \rho_{th} \right| \varphi_- \right\rangle
		=  \left\{
		\begin{array}{lc}
			\ds \!\!\int\limits_{\varphi(0,\bm{x}) = \varphi_-(\bm{x}) }^{\varphi(-i\beta,\bm{x}) = \varphi_+(\bm{x}) } \!\!\!\!\!\!\!\!\!\!\!\!\! \mathcal{D}{\varphi}\ \exp\left( i\int_\I\!d^4x\,\mathcal{L}(x)\right)\\[3ex]
			\ds \hfill \mbox{for\ ``\,$\CpI$\,''} \,, \\[2ex]
			\ds \exp \left( i\sum\limits_{n=0}^\infty \alpha^{\it th}_{12\cdots n} \varphi_1 \varphi_2 \cdots \varphi_n \right)\\
			\ds \hfill \mbox{for\ ``\,$\CpA$\,''} \,,
		\end{array}\right.
	\end{equation}
	into eq.~(\ref{ROOE:GeneratingFunctional}). The argument of the exponential in the lower
	expression is a short-hand notation for
	eqs.~(\ref{ROOE:DensityMatrixElement},\,\ref{ROOE:ExpansionOfDensityMatrixElement}). In the
	following, we show how perturbative Feynman diagrams formulated within the well-known
	$\CpI$ formalism can equivalently be represented within the $\CpA$ formalism.

	\subsection{Thermal time contour $\CpI$}
	
	In this subsection, we briefly review the well-known description of thermal equilibrium based
	on the thermal time contour $\CpI$ in order to establish the notations required later on.
	We stress that, for our purpose, we have to keep the \emph{initial time} finite. For this case,
	the formulation of thermal field theory has been discussed in Refs.~\cite{LeBellac:1996at,%
	Gelis:1994dp,Gelis:1999nx}.
	The free thermal propagator defined on the thermal time contour $\CpI$ is
	\[
		i\GthF^{-1}(x,y) = \left(-\Box_x-m^2\right)\delta_{\CpI}(x-y)\,,
	\]
	for $x^0, y^0\in\CpI$. It may be decomposed into the free thermal statistical propagator
	$\GthFF(x,y)$ and the free thermal spectral function $\GthFR(x,y)$,
	\[ \GthF (x, y) = \GthFF (x, y) - \frac{i}{2} \sgn_{\CpI} (x^0 - y^0) \GthFR(x,y) \;. \]
	The explicit solution of the free equation of motion is
	\begin{eqnarray}\label{ROOE:ThermalClassProp}
		\GthFF(x^0,y^0,\bm{k}) & = & \frac{n_{BE}(\omega_k)+\frac{1}{2}}{\omega_k} \cos\left(\omega_k(x^0-y^0)\right)\,,\nonumber\\
		\GthFR(x^0,y^0,\bm{k}) & = & \frac{1}{\omega_k} \sin\left(\omega_k(x^0-y^0)\right)\,,
	\end{eqnarray}
	for $x^0, y^0\in\CpI$. Here, $n_{BE}(\omega_k)$ is the Bose-Einstein distribution function,
	\[
		n_{BE}(\omega_k) = \frac{1}{e^{\beta\omega_k} - 1}, \qquad
		\omega_k = \sqrt{m^2 + \bm{k}^2}\,.
	\]
	Each of the two time arguments of the propagator can either be real or imaginary, which yields
	four combinations $\GthF^{\C\C}$, $\GthF^{\C\I}$, $\GthF^{\I\C}$, $\GthF^{\I\I}$. These
	appear in perturbative Feynman diagrams which are constructed with the free propagator $\GthF$
	and the classical vertices. In position space, each internal vertex of a Feynman diagram is
	integrated over the thermal time contour $\CpI$. In order to disentangle the contributions
	from the real and the imaginary branch of the time contour, the following Feynman rules are
	defined,
	\[
	\begin{array}{l}
		\begin{array}{cccccc}
			\GthF^{\C\C}(x,y)&=&\feyn{0mm}{0.175}{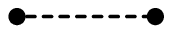} \;, &
			\GthF^{\C\I}(x,y)&=&\feyn{0mm}{0.175}{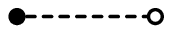} \;, \\[1.6ex]
			\GthF^{\I\I}(x,y)&=&\feyn{0mm}{0.175}{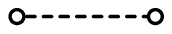} \;, &
			\GthF^{\I\C}(x,y)&=&\feyn{0mm}{0.175}{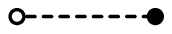} \;,
		\end{array}
	\\[3ex]
		\begin{array}{ccccccccc}
			\ds -i\lambda\!\!\int_{\C}\!\!\!d^4x\!\!&=&\!\!\feyn{-2mm}{0.07}{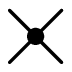} , &
			\ds -i\lambda\!\!\int_{\I}\!\!\!d^4x\!\!&=&\!\!\feyn{-2mm}{0.07}{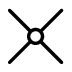} , &
			\ds -i\lambda\!\!\intCI d^4x\!\!&=&\!\!\feyn{-2mm}{0.07}{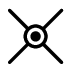} \;.
		\end{array}
	\end{array}
	\]
	Filled circles denote a real time, and empty circles denote an imaginary time.
	As an example, the perturbative setting-sun diagram is considered with propagators attached
	to both external lines, and evaluated for real external times $x^0,y^0\in\C$. Both internal
	vertices are integrated over the two branches $\C$ and $\I$, respectively. Using the
	Feynman rules above, the resulting four contributions can be depicted as
	\begin{eqnarray*}
		\lefteqn{S_0(x,y) \ = \ \feyn{-2.25mm}{0.2}{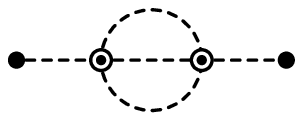}} \\
		& = & \frac{(-i\lambda)^2}{6} \intl_{\CpI} \!\! d^4 u \!\! \intl_{\CpI} \!\! d^4 v \; \GthF(x,u) \GthF^3(u,v) \GthF(v,y) \\
		& = & \feyn{-2mm}{0.185}{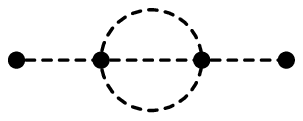}
		      + \feyn{-2mm}{0.185}{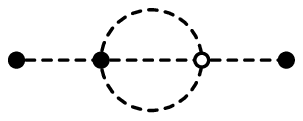}
		      + \feynRot{3.9mm}{0.185}{180}{FreeSettingSunThermalCI}
		      + \feyn{-2mm}{0.185}{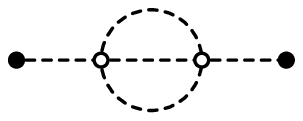} \ .
	\end{eqnarray*}
	We note that  $S_0(x,y)|_{x^0=y^0}$ is time-independent, as expected in thermal
	equilibrium. Nevertheless, the four contributions shown in the last line may
	individually depend on time. However, this time-dependence cancels out in
	their sum, as has been, for example, discussed in Ref.~\cite{Gelis:1994dp}.
	
	\subsection{Closed real-time contour with thermal initial correlations $\CpA$}

	Within the $\CpA$ formalism, all internal vertices of Feynman diagrams are just integrated
	over the closed real-time path $\C$. However, the diagrams may contain non-local effective
	$n$-point vertices. These represent the $n$-point correlations
	$\alpha_n^{\it th}(x_1,\dots,x_n)$ of the thermal initial state. In the following, we show
	how to determine the $\alpha_n^{\it th}$ by a matching procedure employing the
	equivalent $\CpI$ formalism.
	
	Let us consider a Feynman diagram within the $\CpI$ formalism, like for example the
	perturbative setting-sun diagram $S_0(x,y)$. We assume that all time arguments corresponding
	to the external lines are real. It turns out that a single diagram within the $\CpI$ formalism
        is represented by a set of diagrams within the $\CpA$ formalism. Some of these will contain
	non-local effective vertices. Since we are working in the framework of perturbation theory,
	we have to insert approximations to the exact effective vertices. For this purpose, we have to
	determine (i) the topologies of the required diagrams and (ii) the proper approximations for
	the effective vertices.

	In order to do so, we first consider the free thermal propagator evaluated with one imaginary
	and one real time. Using eq.~(\ref{ROOE:ThermalClassProp})
	together with elementary trigonometric addition theorems, it can be written as
	\begin{eqnarray*}
		\GthF^{\I\C}(-i\tau,y^0,\bm{k}) & = & \frac{\GthF^{\I\I}(-i\tau,0,\bm{k})}{\GthF(0,0,\bm{k})}\,\GthFF^{\C\C}(0,y^0,\bm{k})\\
		                                &   & {} + i\partial_\tau\GthF^{\I\I}(-i\tau,0,\bm{k})\,\GthFR^{\C\C}(0,y^0,\bm{k})\,.
	\end{eqnarray*}
	Next, the unequal-time statistical propagator and the spectral function are rewritten as
	\begin{eqnarray*}
		\GthFF^{\C\C}(0,y^0,\bm{k}) & = & \intl_{\C} dz^0 \; \delta_s(z^0)\GthF^{\C\C}(z^0,y^0,\bm{k}) \;, \\
		\GthFR^{\C\C}(0,y^0,\bm{k}) & = & - 2i \intl_{\C} dz^0 \; \delta_a(z^0)\GthF^{\C\C}(z^0,y^0,\bm{k}) \;,
	\end{eqnarray*}
	where
	\begin{eqnarray*}
		\delta_s(z^0) & = & \frac{1}{2}\left(\delta_\C(z^0-0_+) + \delta_\C(z^0-0_-)\right)\,,\nonumber\\
		\delta_a(z^0) & = & \frac{1}{2}\left(\delta_\C(z^0-0_+) - \delta_\C(z^0-0_-)\right)\,.
	\end{eqnarray*}
	Combining these equations, a helpful expression for the free propagator evaluated with
	one imaginary and one real time is obtained,
	\begin{eqnarray}\label{GthCDelta}
		\GthF^{\I\C}(-i\tau,y^0,\bm{k})           & = & \intl_{\C} dt \; \Delta_0(-i\tau,t,\bm{k})\GthF^{\C\C}(t,y^0,\bm{k}) \;, \nonumber \\[1.5ex]
		\feyn{0mm}{0.175}{FreePropagatorIC} \quad & = & \quad\qquad \feyn{-1mm}{0.225}{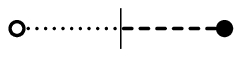}\ .
	\end{eqnarray}
	Here, the \emph{free connection} $\Delta_0(-i\tau,z^0,\bm{k})$ is given by
	\begin{eqnarray}\label{FreeDelta}
		\Delta_0(-i\tau,z^0,\bm{k}) & = &  \Delta_{0}^s(-i\tau,\bm{k})\,\delta_s(z^0) + \Delta_{0}^a(-i\tau,\bm{k})\,\delta_a(z^0) \nonumber \\[1.5ex]
		                            & = & \qquad \feyn{-1mm}{0.15}{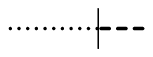}\ ,
	\end{eqnarray}
	where
	\begin{eqnarray}\label{FreeDeltaSA}
		\Delta_{0}^s(-i\tau,\bm{k}) & = & \frac{\GthF^{\I\I}(-i\tau,0,\bm{k})}{\GthF(0,0,\bm{k})}\ , \nonumber\\
		\Delta_{0}^a(-i\tau,\bm{k}) & = & 2\partial_\tau\GthF^{\I\I}(-i\tau,0,\bm{k})\ .
	\end{eqnarray}
	Analogously, the free propagator evaluated with one real and one imaginary time can be
	written as
	\begin{eqnarray}
		\GthF^{\C\I}(y^0,-i\tau,\bm{k})           & = & \intl_{\C} dt \; \GthF^{\C\C}(x^0,t,\bm{k})\Delta_0^T(t,-i\tau,\bm{k}) \nonumber \;, \\[1.5ex]
		\feyn{0mm}{0.175}{FreePropagatorCI} \quad & = & \quad\qquad \feynRot{+3mm}{0.225}{180}{FreePropagatorDelta} \label{GthCDelta.mmm} \;,
	\end{eqnarray}
	$
		\mbox{where} \  \ds\Delta_0^T(z^0,-i\tau,\bm{k}) = \Delta_0(-i\tau,z^0,\bm{k}) =
		\feynRot{+3mm}{0.15}{180}{FreeDelta}\ .
	$
	The connections $\Delta_0$ and $\Delta_0^T$ are attached to an imaginary and a real vertex on
	the left and right sides, respectively. Their Fourier transform into position space is
	\begin{equation}\label{ROOE:FourierTrfOfDelta}
		\Delta_0(v,z) = \int\!\!\! \frac{d^3\!\bm{k}}{(2\pi)^3}\, e^{i\bm{k}(\bm{v}-\bm{z})} \, \Delta_0(v^0,z^0,\bm{k}) \;,
	\end{equation}
	for $v^0\in\I$ and $z^0\in\C$, as well as $\Delta_0^T(z,v)=\Delta_0(v,z)$.
	
	In general, for any thermal diagram on $\CpI$ with $\mathcal{V}$ internal vertices, there are
	$2^\mathcal{V}$ possibilities to combine the integration over $\C$ or $\I$ at each vertex.
	For each of these $2^\mathcal{V}$ contributions, all lines connecting a real and an imaginary
	vertex are replaced using relations~(\ref{GthCDelta}) and (\ref{GthCDelta.mmm}). Thereby the parts
	containing $\I$-integrations are encapsulated into non-local effective vertices. Thus, any thermal
	diagram on $\CpI$ can equivalently be represented by $2^\mathcal{V}$ diagrams on $\C$, which
	contain classical vertices as well as non-local effective vertices.

	For example, the setting-sun diagram with one real and one imaginary vertex can be rewritten
	as
	\begin{equation} \label{FreeSettingSunThermalCI}
		       \feyn{-2.5mm}{0.21}{FreeSettingSunThermalCI}
		=      \feyn{-3.5mm}{0.21}{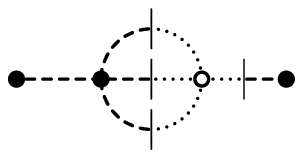}
		\equiv \feyn{-2.75mm}{0.21}{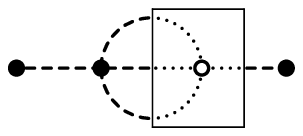}
		\equiv \feyn{-2.5mm}{0.21}{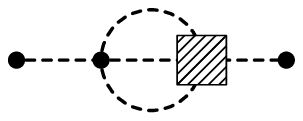}
			\raisebox{+2.75mm}{\hspace*{-4.25mm}\mbox{\tiny\it  th}}
			\raisebox{-2.5mm}{\hspace*{-1.25mm}\tiny  0L}
	\end{equation}
	According to the symbolic notation employed here, the subdiagram containing the imaginary
	vertex, marked by the box, can be encapsulated into an effective non-local 4-point vertex.
	Its structure is determined by the connections $\Delta_0$ and $\Delta_0^T$. This can be seen
	by rewriting the above diagrams in terms of the corresponding formal expressions (only the
	first and last one are given here),
	\begin{eqnarray*}
		\lefteqn{\frac{(-i\lambda)^2}{6} \! \intl_{\C} \! d^4 u \! \intl_{\I} \! d^4 v \;\GthF(x,u)\GthF^3(u,v)\GthF(v,y)}\\
		\!\!  & \equiv & \! \frac{-i\lambda}{6} \!\! \intl_{\C} \!\!  d^4 u \!\!  \intl_{\C} \!\!  d^4 z_{1234} \; \GthF(x,u)\GthF(u,z_1)\GthF(u,z_2) \\
		\!\!  &        & \! {} \times \GthF(u,z_3) \left[ \rule{0mm}{6mm} \alpha_{4,\, 0L}^{th}(z_1,z_2,z_3,z_4) \right]\GthF(z_4,y) \;.
	\end{eqnarray*}
	In the last line, the thermal effective 4-point vertex has been introduced,
	\begin{eqnarray*}
		\alpha_{4, 0L}^{th} (z_1, z_2, z_3, z_4) & = & -i \lambda \intl_{\I} d^4 v \; \Delta_0 (v, z_1) \Delta_0 (v, z_2) \\
		                                         &   & {} \times \Delta_0 (v, z_3) \Delta_0 (v, z_4) \;, \\
		\feyn{-5mm}{0.125}{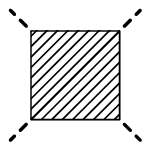}
			\raisebox{+2.75mm}{\hspace*{-1mm}\mbox{\scriptsize\it  th}}
			\raisebox{-3mm}{\hspace*{-2mm}\scriptsize  0L} \qquad
		                                           & = & \qquad \feyn{-5mm}{0.125}{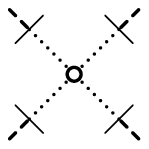} \quad \equiv \quad  \feyn{-5mm}{0.125}{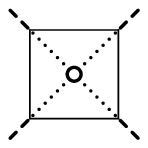} \quad .
	\end{eqnarray*}
	Since the connection $\Delta_0(v,z_i)$ is supported only at the initial time $z_i^0=0_\pm$,
	the effective 4-point vertex vanishes as soon as one of the four real times
	$z^0_1,\dots,z^0_4$ lies beyond the initial time. Thus, the effective 4-point vertex has
	precisely the structure of a non-local effective vertex describing an initial
	correlation. Furthermore, the above 4-point vertex constitutes the leading order contribution
	to the loop expansion of the thermal initial 4-point correlation function
	(see appendix~\ref{QFOOE:sec:ThermalInitialState}).
	
	Diagrams with internal lines connecting {\it two} imaginary vertices contain
	the propagator $\GthF^{\I\I}(-i\tau,-i\tau',\bm{k})$. In order to identify the correct
	effective vertices in this case, the following relation is employed
	\begin{eqnarray}\label{ROOE:GTHFandD0}
		\lefteqn{\GthF^{\I\I}(-i\tau,-i\tau',\bm{k})}\nonumber \\
		& = & D_0(-i\tau,-i\tau',\bm{k}) + \intl_{\C} dw^0 \intl_{\C} dz^0 \; \Delta_0(-i\tau,w^0,\bm{k}) \nonumber \\
		&   & {} \times \GthF^{\C\C}(w^0,z^0,\bm{k})\Delta_0^T(z^0,-i\tau',\bm{k}) \nonumber \\
		& = & D_0(-i\tau,-i\tau'\!,\bm{k}) \nonumber \\
		&   & {} + \Delta_0^s(-i\tau,\bm{k})\GthF^{\C\C}(0,0,\bm{k})\Delta_0^s(-i\tau'\!,\bm{k}) \;,
	\end{eqnarray}
	\[		
		      \feyn{0mm}{0.175}{FreePropagatorII}
		\ = \ \feyn{0mm}{0.175}{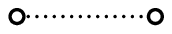}
		\ + \ \feyn{-1mm}{0.38}{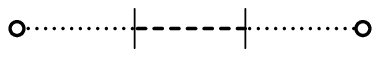}\ .
	\]
	It can be verified by explicit calculation from
	eqs.~(\ref{QFOOE:D0Prop}, \ref{ROOE:ThermalClassProp}, \ref{FreeDelta}).
	Here the propagator $D_0(-i\tau,-i\tau',\bm{k})$, which is defined in
	eq.~(\ref{QFOOE:D0Prop}), is represented by the dotted line. It connects two imaginary times
	and furnishes the perturbative expansion of the thermal initial correlations as
	discussed in appendix~\ref{QFOOE:sec:ThermalInitialState}. Applying the upper relation,
	the setting-sun diagram with two imaginary vertices can be rewritten as
	\[\begin{array}{l}
		          \feyn{-2.5mm}{0.21}{FreeSettingSunThermalII}
		\ = \     \feyn{-2.5mm}{0.21}{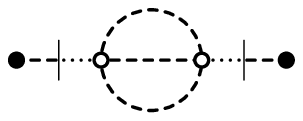}
		\ \equiv\ \feyn{-3.0mm}{0.21}{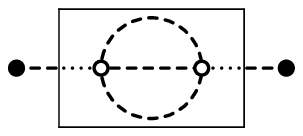}
		\\[3ex]
		=         \feyn{-2.5mm}{0.19}{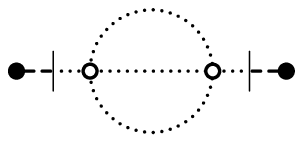}
		+         \feyn{-2.5mm}{0.19}{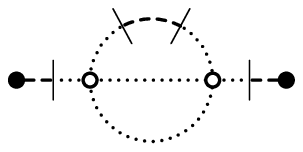}
		+         \feyn{-3.25mm}{0.19}{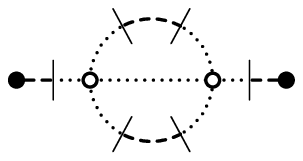}
		+         \feyn{-3.45mm}{0.19}{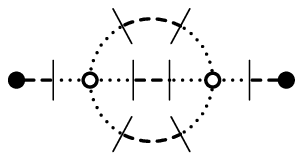}
		\\[3ex]
		=         \feyn{-2.5mm}{0.19}{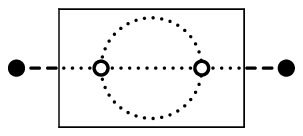}
		+         \feyn{-2.5mm}{0.19}{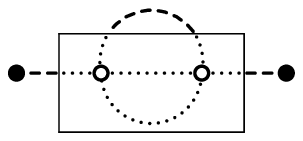}
		+         \feyn{-3.0mm}{0.19}{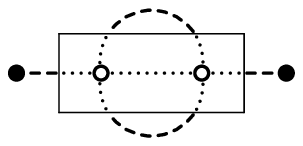}
		+         \feyn{-3.0mm}{0.19}{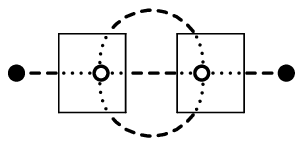}\ .
	\end{array}\]
	In the first step, the propagators connecting real and imaginary vertices were replaced
	using relation~(\ref{GthCDelta}). This already yields an effective non-local two-vertex,
	as indicated in the third diagram in the first line. In order to check that this effective
	non-local two-vertex is indeed composed from the \emph{thermal} initial correlations, the
	three propagators connecting the two imaginary vertices are replaced using
	relation~(\ref{ROOE:GTHFandD0}). Accordingly, the diagram splits into eight terms. These
	can be combined to the four inequivalent contributions shown in the second line%
	\footnote{\label{SymmFactor}%
		Note that the symmetry factors are taken into account properly. For example, the
		symmetry factor of the second diagram in the second line is one third times the
		symmetry factor of the original diagram in the first line. Since there are three
		possibilities to obtain this diagram from the first one, it is obtained with the
		correct prefactor.
	}.
	Finally, the parts containing imaginary vertices and dotted lines can be identified with
	the corresponding contributions to the perturbative expansion of the thermal initial
	correlations discussed in appendix~\ref{QFOOE:sec:ThermalInitialState}. This is represented
	graphically by encapsulating the subdiagrams inside the boxes. In the third line, the first
	diagram thus contains a thermal effective two-point vertex, which itself appears at two-loop
	order in the perturbative expansion of the thermal initial correlations. Similarly, the
	thermal effective four- and six-point vertices contained in the second and third diagram,
	respectively, appear at one- and zero-loop order in the perturbative expansion of the thermal
	initial correlations. The two effective four-point vertices contained in the fourth diagram
	are identical to the one already encountered in eq.~(\ref{FreeSettingSunThermalCI}).
	
	Thus, using the representation (\ref{GthCDelta}) of the free propagator connecting a real and
	an imaginary time and eq.~(\ref{ROOE:GTHFandD0}), any perturbative thermal Feynman diagram
	formulated on the thermal time
	contour $\CpI$ can be related with a set of perturbative Feynman diagrams formulated on the
	closed real-time contour $\C$. Furthermore, the required approximation to the complete thermal
	initial correlations $\alpha^{\it th}_n$ can be explicitly constructed with the help of the
	formalism introduced here. For example, for the perturbative setting-sun diagram, the
	equivalence between ``\,$\CpI$\,'' and ``\,$\CpA$\,'' can, in summary, be written as
	\[\begin{array}{l}
		S_0(x,y) \ = \
		   \feyn{-2.5mm}{0.21}{FreeSettingSunThermal}\ =
		\\[3ex]
		   \feyn{-2mm}{0.19}{FreeSettingSunThermalCC}
		+  \feynRot{4.5mm}{0.19}{180}{FreeSettingSunThermalCIBox}
		+  \feyn{-2.5mm}{0.19}{FreeSettingSunThermalCIBox}
		+  \feyn{-3mm}{0.19}{FreeSettingSunThermalIIThreeDDBox}
		\\[3ex]
		+\,\feyn{-2.35mm}{0.19}{FreeSettingSunThermalIIZeroDDBox}
		+  \feyn{-2.5mm}{0.19}{FreeSettingSunThermalIIOneDDBox}
		+  \feyn{-3.0mm}{0.19}{FreeSettingSunThermalIITwoDDBox}\quad .
	\end{array}\]
	As has been noted before,  $S_0(x,y)|_{x^0=y^0}$ is time-independent. Nevertheless,
	the individual contributions shown above may depend on time, similarly as for the
	$\CpI$-formalism. Since $S_0(x,y)|_{x^0=y^0}$ is time-independent, it is clear that
	the time-dependence of the individual contributions has to cancel when summing them
	up. Thus the time-translation invariance of thermal equilibrium within the $\CpA$-formalism
	is manifestly inherited from the $\CpI$-formalism~\cite{LeBellac:1996at,Gelis:1994dp,%
	Gelis:1999nx} by the matching procedure described here. Since the same
	argument applies for the 2PI case discussed below, we will not repeat it there.

%
\section{Thermal Initial Correlations: 2PI}\label{ROOE:sec:NonPerturbativeThermInitCorr}
%
	
	In this section, the perturbative techniques introduced in the previous section are generalized
	to the nonperturbative 2PI case. This is required since \KB equations are based on the 2PI
	formalism. While the Feynman diagrams shown in the previous section contain the free propagator,
	the diagrams treated here contain the complete propagator, which itself is the solution of a
	self-consistent Schwinger-Dyson equation.

	\subsection{Thermal time contour $\CpI$}
	
	The complete thermal propagator defined on the thermal time path $\CpI$ satisfies the
	self-consistent Schwinger-Dyson equation derived from the 2PI effective action in thermal
	equilibrium,
	\begin{equation}\label{ROOE:GthFull}
		\Gth^{-1}(x,y) = i(\Box_x+m^2)\delta_{\CpI}(x-y) - \Pith(x,y) \,,
	\end{equation}
	where $x^0, y^0\in\CpI$ and $\Pith(x,y)$ is the thermal self-energy. For example, in
	setting-sun approximation it reads
	\[
		\Pith(x,y) = \frac{-i\lambda}{2}\Gth(x,x)\delta_{\CpI}(x-y)
		+ \frac{(-i\lambda)^2}{6}\Gth(x,y)^3 \ .
	\]
	The complete propagator furnishes the expansion of the 2PI effective action in terms of 2PI
	Feynman diagrams. Similar to the perturbative case, the following Feynman rules are defined,
	\[
	\begin{array}{l}
		\begin{array}{cccccc}
			\Gth^{\C\C}(x,y)& = &\feyn{0mm}{0.175}{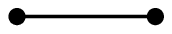} \;, &
			\Gth^{\C\I}(x,y)& = &\feyn{0mm}{0.175}{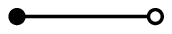} \;, \\[1.6ex]
			\Gth^{\I\I}(x,y)& = &\feyn{0mm}{0.175}{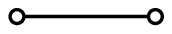} \;, &
			\Gth^{\I\C}(x,y)& = &\feyn{0mm}{0.175}{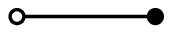} \;,
		\end{array}
	\\[3ex]
		\begin{array}{ccccccccc}
			\ds -i\lambda\!\!\int_{\C}\!\!\!d^4x\!\! & = & \!\!\feyn{-2mm}{0.07}{VertexC} , &
			\ds -i\lambda\!\!\int_{\I}\!\!\!d^4x\!\! & = & \!\!\feyn{-2mm}{0.07}{VertexI} , &
			\ds -i\lambda\!\!\intCI d^4x\!\!     & = & \!\!\feyn{-2mm}{0.07}{VertexCI} \;.
		\end{array}
	\end{array}
	\]
	Accordingly, diagrams containing the complete propagator can be decomposed in analogy to
	the perturbative case. For example,
	\begin{eqnarray*}
		\lefteqn{S(x,y) \ = \ \feyn{-2.25mm}{0.2}{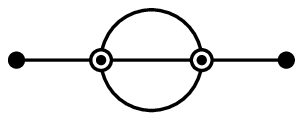}} \\
		& = & \frac{(-i\lambda)^2}{6} \intl_{\CpI} d^4 u \intl_{\CpI} d^4 v \; \Gth(x,u) \Gth^3(u,v) \Gth(v,y) \\
		& = & \feyn{-2mm}{0.185}{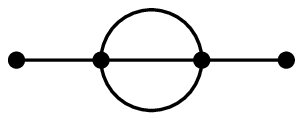}
		      + \feyn{-2mm}{0.185}{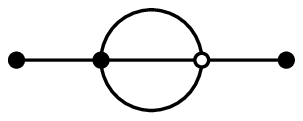}
		      + \feynRot{3.9mm}{0.185}{180}{SettingSunThermalCI}
		      + \feyn{-2mm}{0.185}{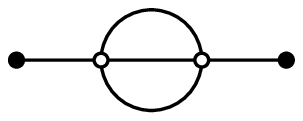} \ .
	\end{eqnarray*}

	\subsection{Closed real-time contour with thermal initial correlations $\CpA$}
	
	As for the perturbative case, we will now use the equivalence between $\CpI$ and
	$\CpA$ in order to infer the proper thermal initial correlations
	$\alpha_n^{\it th}(x_1,\dots,x_n)$ within the 2PI framework.
	
	In order to disentangle the contributions from the real and the imaginary branch of the
	thermal time contour, a generalization of eq.~(\ref{GthCDelta}) to the 2PI case
	is required. Due to the nonperturbative nature of the 2PI formalism, this generalization
	is non-trivial and requires a somewhat lengthy calculation that can be found in
	appendix~\ref{QFOOE:sec:FullConnection}.
	The most important result is that the complete propagator connecting imaginary and real
	times can be decomposed into a convolution of a \emph{complete connection}
	$\Delta(-i\tau,z^0,\bm{k})$ and the complete real-real propagator,
	\begin{eqnarray}\label{Delta}
		      \Gth^{\I\C}(-i\tau,y^0,\bm{k}) 
		& = & \intl_{\C} d t \; \Delta(-i\tau,t,\bm{k})\Gth^{\C\C}(t,y^0,\bm{k}) \;, \nonumber \\
		\feyn{0mm}{0.175}{PropagatorIC} \quad & = & \qquad \feyn{-1mm}{0.22}{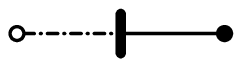} \quad .
	\end{eqnarray}
	Here, the complete connection is given by
	\begin{eqnarray}
		      \Delta(-i\tau,z^0,\bm{k}) 
		& = & \Delta^s(-i\tau,\bm{k})\delta_s(z^0) + \Delta^a(-i\tau,\bm{k})\delta_a(z^0) \nonumber \\
		&   & {} + \intl_{\I} d v^0 \; D(-i\tau,v^0,\bm{k})\Pith^{nl}(v^0,z^0,\bm{k}) \;, \nonumber\\
		      \feyn{-1.75mm}{0.175}{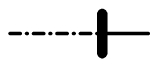}
		& = & \quad\feyn{-1.5mm}{0.175}{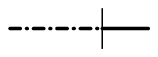}\quad + \quad\feyn{-3mm}{0.2}{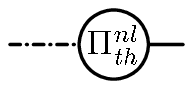} \ .\label{ROOE:CompleteDelta}
	\end{eqnarray}
	While the first line is already known from the perturbative case, the second line is a new
	contribution. It contains the non-local part of the thermal self-energy. The quantities
	$\Delta^s$, $\Delta^a$ and $D$ are straightforward generalizations of their perturbative
	counterparts $\Delta^s_0$, $\Delta^a_0$ and $D_0$ (see eqs.~(\ref{FreeDeltaSA},\ref{ROOE:DeltaSA})
	and eqs.~(\ref{ROOE:GTHFandD0},\ref{ROOE:Dtautau})).

	In contrast to the perturbative case, eq.~(\ref{ROOE:CompleteDelta}) is an implicit equation
	for the complete connection. For example, for the 2PI setting-sun approximation, it has the
	form,
	\[
		\feyn{-1.5mm}{0.175}{Delta} \quad = \quad \feyn{-1.5mm}{0.175}{DeltaM} \quad
		+ \quad \feyn{-5.5mm}{0.25}{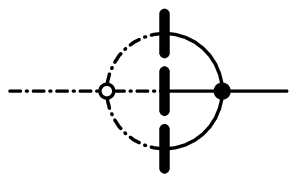} \quad .
	\]
	Equation~(\ref{ROOE:CompleteDelta}) can be solved iteratively,
	\begin{eqnarray}
		      \Delta^{(0)}
		& = & \Delta^s\delta_s(z^0) + \Delta^a\delta_a(z^0) \ = \  \feyn{-1.5mm}{0.175}{DeltaM} \;, \label{ROOE:DeltaIteration} \\
		      \Delta^{(k+1)}
		& = & \Delta^s\delta_s(z^0) + \Delta^a\delta_a(z^0) + \intl_{\I} d v^0 \; D \Pith^{nl} \big|_{\Delta^{(k)}} \;.  \nonumber
	\end{eqnarray}
	For example, for the setting-sun approximation, the first three steps of this iteration are
	\begin{eqnarray*}
		\lefteqn{\feyn{-1.7mm}{0.175}{Delta} \ = \ \feyn{-1.5mm}{0.175}{DeltaM} \ + \ \feyn{-5.25mm}{0.25}{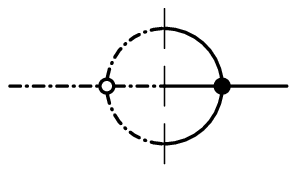}}\\
		& + & \feyn{-7.0mm}{0.25}{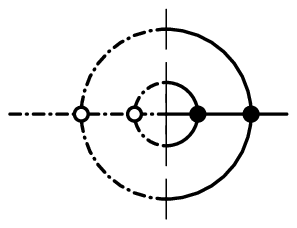}  \ + \  \feyn{-9.0mm}{0.25}{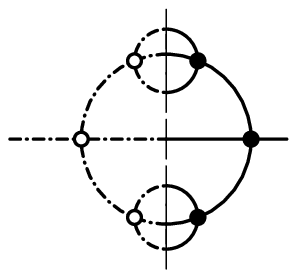}  \ + \  \feyn{-9.0mm}{0.25}{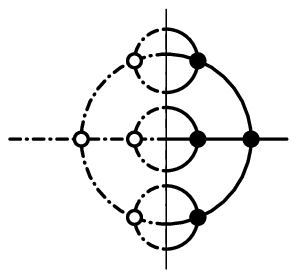}\\
		& + & \dots \;.
	\end{eqnarray*}
	The first line represents the zeroth step and the first step, and the second line shows
	all diagrams contributing at the second step.
	All diagrams are generated with the correct symmetry factors.

	On the other hand, the nonperturbative generalization of eq.~(\ref{ROOE:GTHFandD0}) reads
	\begin{eqnarray}
		      \lefteqn{\Gth^{\I\I}(-i\tau,-i\tau',\bm{k})} \nonumber \\
		& = & D(-i\tau,-i\tau',\bm{k}) + \intl_{\C} dw^0 \intl_{\C} dz^0 \; \Delta(-i\tau,w^0,\bm{k}) \nonumber \\
		&   & {} \times \Gth^{\C\C}(w^0,z^0,\bm{k}) \Delta^T(z^0,-i\tau',\bm{k}) \nonumber \\
 		& = & D(-i\tau,-i\tau',\bm{k}) \nonumber \\
		&   & {} + \Delta^s(-i\tau,\bm{k}) \Gth(0,0,\bm{k}) \Delta^s(-i\tau',\bm{k}) \;. \label{ROOE:Dtautau.mmm}
	\end{eqnarray}
	The derivation of this equation is also shown in appendix \ref{QFOOE:sec:FullConnection}.
	Similar to the perturbative case, the formalism established above can be used to relate any
	Feynman diagram formulated on the thermal time path (``$\CpI$'') with a set of Feynman
	diagrams formulated on the closed real-time path $\C$ containing non-local effective
	vertices representing the thermal initial correlations (``$\CpA$'').
	This is accomplished by three steps:
	\begin{enumerate}
		\item First, the contour integrations over the thermal time path $\CpI$ associated
			with internal vertices are split into two integrations over $\C$ and $\I$.
			A diagram with $\mathcal{V}$ vertices is thus decomposed into 
			$2^{\mathcal{V}}$ contributions.
		\item Next, all internal propagator lines connecting a real and an imaginary time
			are replaced using eq.~(\ref{Delta}). Additionally, the internal
			propagator lines connecting two imaginary times are replaced according to
			eq.~(\ref{ROOE:Dtautau.mmm}). The parts containing imaginary times are
			encapsulated, which can be visualized by joining the \emph{complete
                        connections} to boxes surrounding the imaginary vertices.
		\item Eventually, the iterative solution of eq.~(\ref{ROOE:CompleteDelta}) for the
			complete connection is inserted. Each resulting contribution can be
			identified as a diagram formulated on the closed real-time path $\C$
			containing non-local effective vertices $\alpha_n$. The latter are
			constructed explicitly, as appropriate for the underlying 2PI approximation.
	\end{enumerate}
	The first two steps are analogous to the perturbative case, with complete propagators and
	connections instead of free ones. The third step is special for the nonperturbative case.
	It results in contributions which contain non-local effective vertices $\alpha_n$
	of arbitrarily high order $n$. These take into account the infinite sequence of thermal
	initial $n$-point correlations, which are present due to the underlying nonperturbative
	approximation.
	
	For example, for the nonperturbative setting sun diagram, step one and two can be written as
	\begin{eqnarray*}
		S(x,y) & = & \feyn{-2.25mm}{0.19}{SettingSunThermalCC}
			     + \feyn{-3.0mm}{0.19}{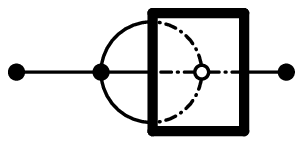}
			     + \feynRot{4.7mm}{0.19}{180}{SettingSunThermalCIThickBox} \\
		       &   & {} + \feyn{-3.0mm}{0.19}{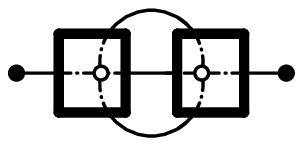}
			     + \feyn{-3.0mm}{0.19}{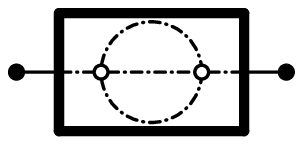}
			     + \feyn{-3.0mm}{0.19}{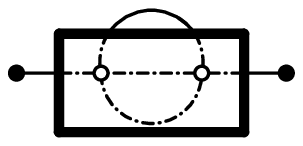} \\
		       &   & {} + \feyn{-3.0mm}{0.19}{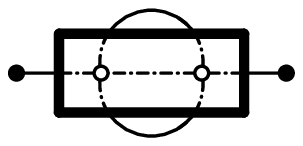} \;.
	\end{eqnarray*}
	For the second diagram, the third step can be written as
	\begin{eqnarray*}
		\lefteqn{
			\feyn{-3.25mm}{0.21}{SettingSunThermalCIThickBox} \ = \
			\feyn{-3.0mm}{0.21}{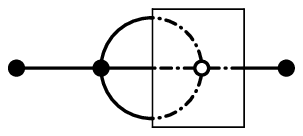}
			+ \feyn{-4.0mm}{0.21}{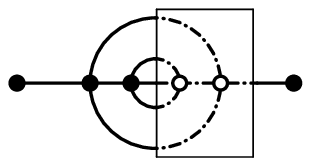}
		} \\
		&  & {} + \feyn{-5.0mm}{0.21}{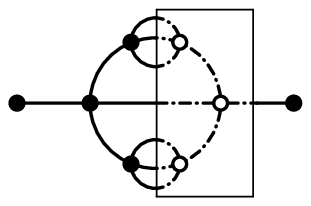}
		     + \feyn{-5.0mm}{0.21}{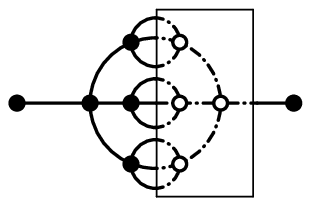}
		     + \feyn{-3.5mm}{0.21}{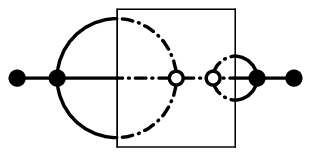} \\
		&  & {} + \feyn{-3.5mm}{0.21}{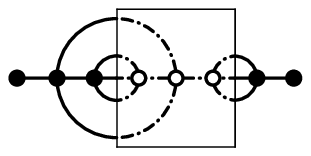}
		     + \feyn{-4.75mm}{0.21}{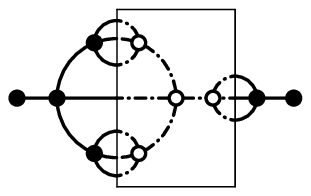}
		     + \feyn{-4.75mm}{0.21}{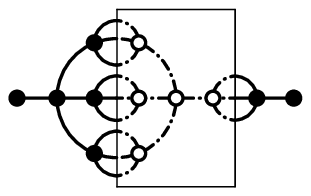}
		     + \ldots \;.
	\end{eqnarray*}
	The first diagram on the right-hand side is obtained by inserting the zeroth iteration for the
	four complete connections. The other diagrams are obtained by inserting the first iteration.
	Note that all diagrams shown above are generated with correct symmetry factors.
	
	Each of the boxes with thin lines represents a non-local effective vertex, encoding the
	correlations of the initial state. A thin box that is attached to $n$ propagator lines
	represents a contribution to the thermal initial $n$-point correlation function
	$\alpha_n^{\it th}(x_1,\dots,x_n)$.
	For example, the leading contributions to the thermal initial $4$- and $6$-point correlations
	are given by
	\begin{eqnarray*}
		\lefteqn{\alpha_{4, 0L}^{th, 2PI}(z_1,\ldots,z_4)} \\
		& = & -i\lambda \! \intl_{\I} \!\! d^4 v \; \Delta^{(0)}(v,z_1)\Delta^{(0)}(v,z_2)\Delta^{(0)}(v,z_3)\Delta^{(0)}(v,z_4)
	\end{eqnarray*}
	\begin{eqnarray}\label{ROOE:ThermalAlpha4}
		\feyn{-4mm}{0.1}{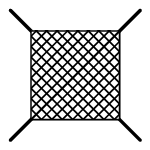}
			\raisebox{+2mm}{\hspace*{-1mm} \mbox{\scriptsize{\it  th},\,2PI}}
			\raisebox{-3mm}{\hspace*{-8mm}\scriptsize  0L}\qquad
		& = & \quad \feyn{-4mm}{0.1}{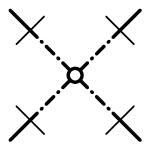}
		      \quad \equiv \quad \feyn{-3.75mm}{0.1}{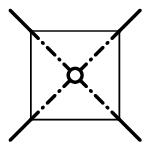}\ ,
	\end{eqnarray}
	and
	\begin{eqnarray*}
		\lefteqn{\alpha_{6, 0L}^{th, 2PI}(z_1,\ldots,z_6)} \\
		& = & (-i\lambda)^2 \intl_{\I} d^4 v \! \intl_{\I} d^4 w \; \Delta^{(0)}(v,z_1)\Delta^{(0)}(v,z_2)\Delta^{(0)}(v,z_3) \\
		&   & {} \times D(v,w)\Delta^{(0)}(w,z_4)\Delta^{(0)}(w,z_5)\Delta^{(0)}(w,z_6) \;,
	\end{eqnarray*}
	\begin{eqnarray}\label{ROOE:ThermalAlpha6}
		\feynRot{-3.5mm}{0.1}{90}{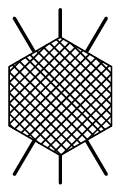}
			\raisebox{+2mm}{\hspace*{-3.5mm} \mbox{\scriptsize{\it  th},\,2PI}}
			\raisebox{-2mm}{\hspace*{-7mm}\scriptsize  0L}\quad
		& = & \ \feyn{-3.5mm}{0.175}{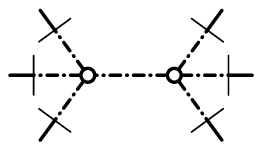} \  \equiv \   \feyn{-3.75mm}{0.175}{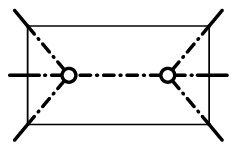}\ .
	\end{eqnarray}
	Note that these contributions are nonperturbative approximations of the exact thermal
	initial correlations, since they involve the complete thermal propagator.

%
\section{\KB Equations with Thermal Initial Correlations}\label{ROOE:sec:KBEThermalCTP}
%

	In this section, \KB equations that can describe thermal equilibrium on the closed real-time
	path $\C$ with finite initial time $t_{init}=0$ are derived. This requires to
	take into account non-Gaussian correlations of the initial (thermal) state of the system.
	These thermal initial correlations have to be determined in accordance with the
	nonperturbative 2PI formalism underlying the \KB equations. Therefore, the techniques
	developed in the previous section are combined with the \KB equations for non-Gaussian
	initial states derived in section~\ref{ROOE:sec:EAaKBE}. As before, the main idea is
	to match the \KB equations describing thermal equilibrium within the $\CpA$-formalism
	on the one hand with equivalent evolution equations obtained from the 2PI effective action
	within the $\CpI$-formalism on the other hand.
	
	On the one hand, we use that the \KB equation for a thermal initial state is a special case of
	the \KB equation for a non-Gaussian initial state (see eq.~(\ref{ROOE:KBENonGauss})), which
	has the form
	\begin{eqnarray}
		\lefteqn{\left(\partial_{x^0}^2+\bm{k}^2+M^2_{\it th}\right)\Gth(x^0,y^0,\bm{k}) = -i \delta_{\C} (x^0 - y^0)} \nonumber \\
		& & {} -i \intl_{\C} dz^0 \bigg[ \Pi^{\it G}_{\it th,nl}(x^0,z^0,\bm{k}) + \Pi^{\it nG}_{\it th,nl}(x^0,z^0,\bm{k}) \nonumber \\
		& & {} + i\Pi_{{\it th},\lambda\alpha}(x^0,z^0,\bm{k})\bigg]\Gth(z^0,y^0,\bm{k}) \;. \label{ROOE:ThermalKBEonCTP}
	\end{eqnarray}
	Here $\Pi^{\it G}_{\it th,nl}(x^0,z^0,\bm{k})$ and $\Pi^{\it nG}_{\it th,nl}(x^0,z^0,\bm{k})$
	denote the Gaussian- and non-Gaussian parts of the non-local self-energy, respectively.
	Furthermore,
	\begin{eqnarray*}
		\lefteqn{\Pi_{{\it th},\lambda\alpha}(x^0,z^0,\bm{k})} \\
		& = & \Pi_{th,\lambda\alpha,F}(x^0,\bm{k})\delta_s(z^0) - \frac{i}{2}\Pi_{th,\lambda\alpha,\rho}(x^0,\bm{k})\delta_a(z^0)
	\end{eqnarray*}
	denotes the contribution from the non-Gaussian initial correlations that is supported at the
	initial time surface $z^0=0$ (see section~\ref{ROOE:sec:EAaKBE}).
	
	On the other hand, the equation of motion of the complete thermal propagator based on the
	thermal time contour (``$\CpI$'') evaluated for $x^0,y^0\in\C$ is
	\begin{eqnarray}
		\lefteqn{\left(\partial_{x^0}^2+\bm{k}^2+M^2_{th}\right)\Gth(x^0,y^0,\bm{k}) = -i \delta_{\CpI} (x^0 - y^0)} \nonumber \\
		& & {} - i \intl_{\CpI} dz^0 \; \Pith^{nl}(x^0,z^0,\bm{k})\Gth(z^0,y^0,\bm{k}) \;. \qquad\qquad \label{ROOE:EoMofThermalPropagator}
	\end{eqnarray}
	Of course, the thermal propagator is time-translation invariant and thus only depends on
	the time difference $x^0 - y^0$. The upper notation is chosen for convenience, in order
	to simplify the comparison with the corresponding \KB equations.

	Since the two formulations of thermal equilibrium are equivalent, the solutions of
	the equation of motion~(\ref{ROOE:EoMofThermalPropagator}) evaluated for $x^0,y^0\in\C$ as well
	as the \KB equation~(\ref{ROOE:ThermalKBEonCTP}) for a thermal initial state have to agree.
	Now we will use this equivalence in order to determine the non-Gaussian parts of the
	thermal self-energy appearing on right-hand side of the \KB equation.

	In order to do so, we start from eq.~(\ref{ROOE:EoMofThermalPropagator}) and split the
	contour integration on the right-hand side into one integration over $\C$ and one over $\I$.
	We note that the resulting two contributions are in general not time-translation invariant,
	but their sum certainly is.
	The integral over the imaginary-time contour $\I$ can be rewritten using the complete
	connection~(\ref{ROOE:CompleteDelta}),
	\begin{eqnarray*}
		\lefteqn{\intl_{\I} dz^0 \; \Pith^{nl}(x^0,z^0,\bm{k}) \Gth(z^0,y^0,\bm{k})} \\
		& = & \intl_{\I} \! dv^0 \; \Pith^{nl}(x^0,v^0,\bm{k}) \intl_{\C} \! dz^0 \; \Delta(v^0,z^0,\bm{k}) \Gth(z^0,y^0,\bm{k}) \\
		& = & \intl_{\C} dz^0 \intl_{\I} dv^0 \; \Pith^{nl}(x^0,v^0,\bm{k}) \bigg( \Delta^{(0)}(v^0,z^0,\bm{k}) \\
		&   & {} + \intl_{\I} \! dw^0 \; D(v^0,w^0,\bm{k}) \Pith^{nl}(w^0,z^0,\bm{k}) \bigg) \Gth(z^0,y^0,\bm{k}) \;.
	\end{eqnarray*}
	Inserting this into the equation of motion~(\ref{ROOE:EoMofThermalPropagator}), it takes
	precisely the form of the \KB equation~(\ref{ROOE:ThermalKBEonCTP}) for a non-Gaussian
	initial state. By comparison, the non-Gaussian contributions to the self-energy for the
	thermal initial state can be inferred,
	\begin{equation} \label{ROOE:ThermalSelfEnergyCTP}
		\Pi^G_{th, nl} (x^0, z^0, \bm{k}) = \left. \Pith^{nl} (x^0, z^0, \bm{k}) \right|_{x^0,z^0\in\C} \;,
	\end{equation}
	\begin{eqnarray*}
		\lefteqn{\Pi^{nG}_{th,nl} (x^0, z^0, \bm{k}) = \intl_{\I} dv^0 \intl_{\I} dw^0 \; \Pith^{nl} (x^0, v^0, \bm{k})} \\
		&   & {} \times \left. D(v^0, w^0, \bm{k}) \Pith^{nl} (w^0, z^0, \bm{k}) \right|_{x^0,z^0\in\C} \;,
	\end{eqnarray*}
	\begin{eqnarray*}
		i \Pi_{th, \lambda\alpha} (x^0, z^0, \bm{k}) & = & \intl_{\I} dv^0 \; \Pith^{nl}(x^0, v^0, \bm{k}) \\
		                                             &   & {} \times \left. \Delta^{(0)}(v^0, z^0, \bm{k}) \right|_{x^0,z^0\in\C} \;.
	\end{eqnarray*}
	For the setting-sun approximation, the steps listed above leading from the formulation of the
	Schwinger-Dyson equation on the thermal time path (``$\CpI$'') to the formulation on the closed
	real-time path with thermal initial correlations (``$\CpA$'') are
	\begin{eqnarray*}
		      \feyn{-2.5mm}{0.1725}{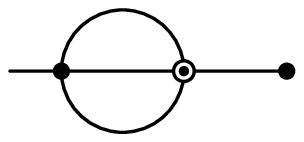}
		& = & \feyn{-2.5mm}{0.1725}{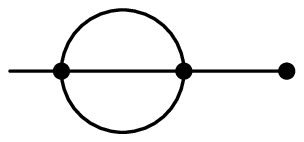}
		      + \feyn{-2.5mm}{0.1725}{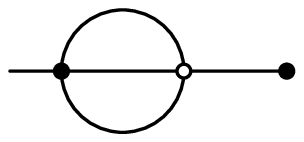} \\
		& = & \feyn{-2.5mm}{0.1725}{SettingSunPropAndDelta_2}
		      + \feyn{-3.75mm}{0.1725}{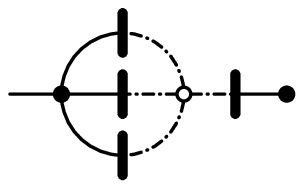} \\
		& = & \feyn{-2.5mm}{0.1725}{SettingSunPropAndDelta_2}
		      + \feyn{-3.75mm}{0.1725}{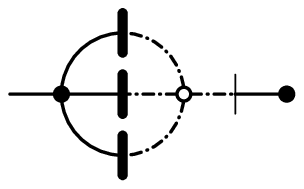}
		      + \feyn{-3.7mm}{0.3}{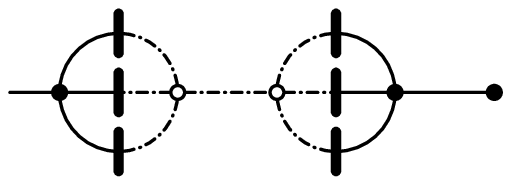}
	\end{eqnarray*}
	Thus, the Gaussian and non-Gaussian contributions to the self-energy in setting-sun
	approximation for a thermal initial state are given by
	\begin{eqnarray}\label{ROOE:ThermalSelfEnergyCTPSettingSun}
		\Pi^{\it G}_{\it th,nl}(x^0,z^0,\bm{k})       & = & \feyn{-4.0mm}{0.18}{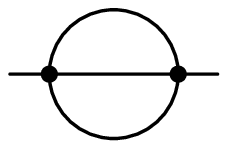} \quad , \nonumber \\
		\Pi^{\it nG}_{\it th,nl}(x^0,z^0,\bm{k})      & = & \feyn{-5.75mm}{0.375}{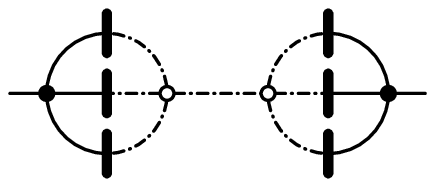} \quad , \\
		i\Pi_{{\it th},\lambda\alpha}(x^0,z^0,\bm{k}) & = & \feyn{-5.6mm}{0.23}{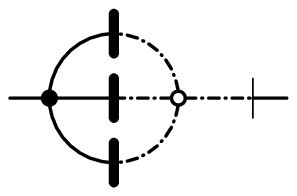} \quad .\nonumber
	\end{eqnarray}
	We stress that the propagator determined by the \KB equations for a thermal initial
	state is time-translation invariant, as required. The matching procedure guarantees that
	this property is directly inherited from the $\CpI$ formalism. Nevertheless, the
	contributions to the self-energy derived above are not, and cannot be, time-translation
	invariant individually (compare with the discussion in
	section~\ref{ROOE:sec:PerturbativeThermInitCorr} and Refs.~\cite{LeBellac:1996at,%
	Gelis:1994dp,Gelis:1999nx}).

	Finally, in order to explicitly obtain the thermal initial correlations that are appropriate
	for a specific 2PI approximation, the iterative expansion~(\ref{ROOE:DeltaIteration}) of the
	complete connection has to be inserted. This yields a series expansion of the non-Gaussian
	self-energies,
	\[
		\Pi_{{\it th},\lambda\alpha} = \sum_{k=0 }^\infty \Pi^{(k)}_{{\it th},\lambda\alpha}\,,\qquad
		\Pi^{\it nG}_{\it th,nl}     = \sum_{k=0 }^\infty \Pi^{(k),{\it\, nG}}_{\it th,nl}\,,
	\]
	where
	\begin{eqnarray}
		\Pi^{(0)}_{{\it th},\lambda\alpha} & = & \left.\Pi_{th,\lambda\alpha}\rule{0mm}{3mm}\right|_{\Delta^{(0)}}\ ,\nonumber \\
		\Pi^{(k)}_{{\it th},\lambda\alpha} & = & \left.\Pi_{th,\lambda\alpha}\rule{0mm}{3mm}\right|_{\Delta^{(k)}} - \Pi^{(k-1)}_{{\it th},\lambda\alpha}\ ,\nonumber
	\end{eqnarray}
	and analogously for $\Pi^{\it\, nG}_{\it th,nl}$.
	
	For example, in setting-sun approximation, the thermal initial correlations obtained from
	inserting the zeroth, first and second iteration of the complete connection are
	\begin{eqnarray}\label{ROOE:PiThLambdaAlpha}
		i\Pi^{(0)}_{{\it th},\lambda\alpha} & = & \feyn{-3.75mm}{0.22}{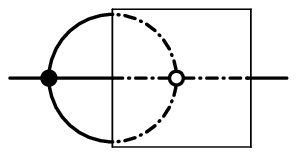}
		                                          \ = \ \feyn{-3.8mm}{0.21}{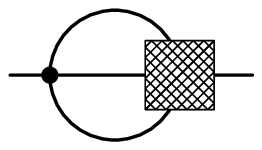}
		                                          \raisebox{+3mm}{\hspace*{-3mm} \tiny\mbox{\it th},\,2PI}
		                                          \raisebox{-2mm}{\hspace*{-8mm} {\tiny  0L}} \ ,\\[1.5ex]
		i\Pi^{(1)}_{{\it th},\lambda\alpha} & = & \feyn{-4.85mm}{0.22}{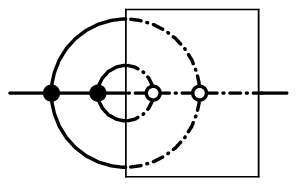}
		                                          + \feyn{-6.25mm}{0.22}{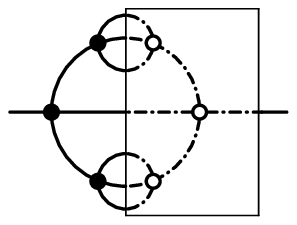}
		                                          + \feyn{-6.25mm}{0.22}{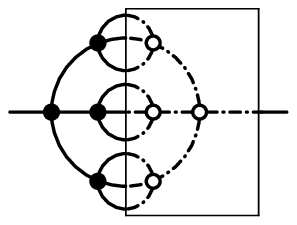} \ , \nonumber \\[1.5ex]
		i\Pi^{(2)}_{{\it th},\lambda\alpha} & = & \feyn{-7.35mm}{0.28}{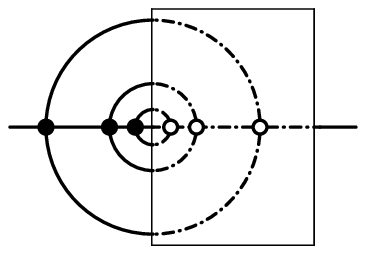}
		                                          + \ldots
		                                          + \feyn{-9.75mm}{0.28}{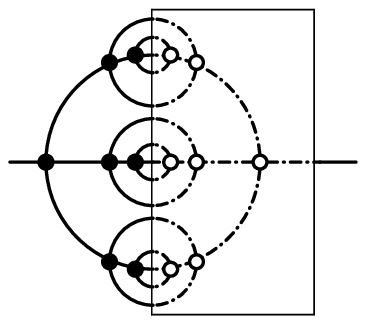} \ . \nonumber
	\end{eqnarray}
	The zeroth contribution contains the thermal non-local effective $4$-point
	vertex~(\ref{ROOE:ThermalAlpha4}). The first iteration yields three diagrams with thermal
	effective $6$-, $8$-, and $10$-point vertices, and the second iteration yields six
	contributions with thermal effective $8$-\nolinebreak, $10$-\nolinebreak, $12$-
	(two diagrams), \parbox[b]{7mm}{$14$-,} and $16$-point vertices. The smallest and largest
	are shown in the last line of eq.~(\ref{ROOE:PiThLambdaAlpha}).
	The expansion of $\Pi^{\it nG}_{\it th,nl}$ contains thermal non-local effective vertices
	of order six and higher,
	\begin{eqnarray*}
		\Pi^{(0), nG}_{th, nl} & = & \feyn{-4.2mm}{0.35}{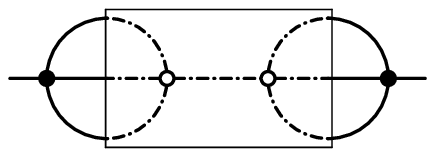}
		                             \ = \ \feynRot{-3.75mm}{0.11}{90}{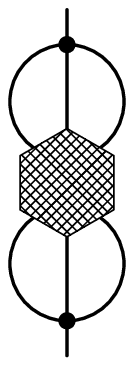}
		                             \raisebox{+5.25mm}{\hspace*{-18mm} \tiny\mbox{\tiny\it th},2PI}
		                             \raisebox{-5mm}{\hspace*{-4mm} {\tiny  0L}} \qquad\quad\ ,\\
		\Pi^{(1), nG}_{th, nl} & = & \feyn{-5.2mm}{0.35}{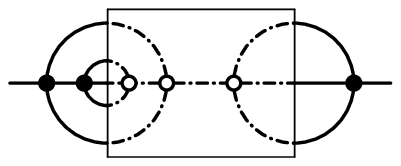}
		                             + \feyn{-5.2mm}{0.35}{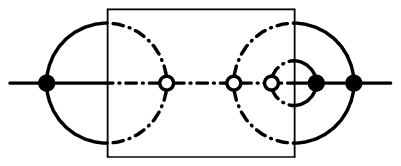} \\
		                       &   & {} + \ldots
		                             + \feyn{-6.25mm}{0.35}{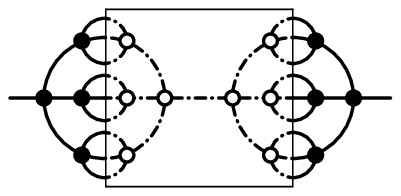} \ .
	\end{eqnarray*}
	The zeroth contribution contains the thermal non-local effective $6$-point
	vertex~(\ref{ROOE:ThermalAlpha6}). The first contribution contains $15$ diagrams with
	thermal effective vertices of order $8$ to $18$.
	\begin{table}
		\begin{center}
			$ \Pi_{{\it th},\lambda\alpha}(x^0,z^0,\bm{k}) $
		\end{center}
		\begin{tabular}{c|cccccccccccccccccc}
			  &     4    &    6     &    8     &    10    &    12    &    14    &    16    & $\cdots$ &    22    & $\cdots$ &    28    & $\cdots$ &     34    & $\cdots$\\
			\hline
			0 & $\times$ \\
			1 &          & $\times$ & $\times$ & $\times$ \\
			2 &          &          & $\times$ & $\times$ & $\times$ & $\times$ & $\times$\\
			3 &          &          &          & $\times$ & $\times$ & $\times$ & $\times$ & $\cdots$ & $\times$\\
			4 &          &          &          &          & $\times$ & $\times$ & $\times$ & $\cdots$ & $\times$ & $\cdots$ & $\times$\\
			5 &          &          &          &          &          & $\times$ & $\times$ & $\cdots$ & $\times$ & $\cdots$ & $\times$ & $\cdots$ & $\times$\\
			$\vdots$
		\end{tabular}
		\begin{center}
			$ \Pi^{\it nG}_{\it th,nl}(x^0,z^0,\bm{k}) $
		\end{center}
		\begin{tabular}{c|cccccccccccccccccc}
			  &     4    &    6     &    8     &    10    &    12    &    14    &    16    &    18    & $\cdots$ &    30    & $\cdots$ &    42    & $\cdots$ &    54    & $\cdots$ &    66    & $\cdots$\\
			\hline
			0 &          & $\times$ \\
			1 &          &          & $\times$ & $\times$ & $\times$ & $\times$ & $\times$ & $\times$\\
			2 &          &          &          & $\times$ & $\times$ & $\times$ & $\times$ & $\times$ & $\cdots$ & $\times$\\
			3 &          &          &          &          & $\times$ & $\times$ & $\times$ & $\times$ & $\cdots$ & $\times$ & $\cdots$ & $\times$\\
			4 &          &          &          &          &          & $\times$ & $\times$ & $\times$ & $\cdots$ & $\times$ & $\cdots$ & $\times$ & $\cdots$ & $\times$\\
			5 &          &          &          &          &          &          & $\times$ & $\times$ & $\cdots$ & $\times$ & $\cdots$ & $\times$ & $\cdots$ & $\times$ & $\cdots$ & $\times$\\
			$\vdots$
		\end{tabular}
		\caption{\label{ROOE:tab:ThermalInitCorr}\captionstyle%
			Thermal initial correlations in 2PI setting-sun approximation. The column
			number is the order $n = 4, 6, \dots$ of the thermal initial $n$-point
			correlation. The row number $k = 0, 1, \dots$ shows which initial correlations
			contribute to $\Pi^{(k)}_{{\it th},\lambda\alpha}$ (upper table) and
			$\Pi^{(k),{\it\, nG}}_{\it th,nl}$ (lower table), respectively. Due to the
			$Z_2$-symmetry, only even correlations are non-zero.
		}
	\end{table}
	The order of the thermal initial correlations appearing up to the fifth contribution in
	setting-sun approximation are shown in table~\ref{ROOE:tab:ThermalInitCorr}.

	Thus, vacuum or thermal initial states entail an infinite hierarchy of initial $n$-point
	correlation functions. In setting-sun approximation, the non-Gaussian initial correlation of
	lowest order is the $4$-point correlation given in the first line of
	eq.~(\ref{ROOE:PiThLambdaAlpha}). Its contribution to the \KB equations is
	\begin{eqnarray*}
		i\Pi^{(0)}_{th, \lambda\alpha} (x,z) & = & \frac{-i\lambda}{6} \intl_{\C} d^4 x_{123} \; \Gth(x,x_1) \Gth(x,x_2) \\
						     &   & {} \times \Gth(x,x_3) i \alpha^{th, 2PI}_{4, 0L} (x_1,x_2,x_3,z) \;.
	\end{eqnarray*}
	In the limit $x^0, y^0\rightarrow 0$, only the thermal initial $4$-point correlation
	contributes to the right-hand side of the \KB equations~(\ref{ROOE:ThermalKBEonCTP}):
	\begin{eqnarray}
		\lefteqn{\intl_{\C} dz^0 \left[ \Pi^G_{th, nl} + \Pi^{nG}_{th, nl} + i \Pi_{th, \lambda\alpha} \right] (x^0, z^0, \bm{k}) \Gth(z^0, y^0, \bm{k})} \nonumber \\
		& \to & \intl_{\C} dz^0 \; i \Pi^{(0)}_{th, \lambda\alpha} (x^0, z^0, \bm{k}) \Gth(z^0, y^0, \bm{k}) \;. \hspace{20mm} \label{ROOE:PiThAtInitialTime}
	\end{eqnarray}
	The reason is that all other contributions contain at least one internal classical
	vertex. However, these vertices are accompanied by memory integrals. Due to the
	structure of the closed real time-path, memory integrals vanish when all external time
	arguments approach the initial time.
	
	Altogether, \KB equations that describe thermal equilibrium have been derived on the closed
	real-time path with finite initial time. This requires to take into account non-Gaussian
	$n$-point correlations of the initial state, which enter explicitly on the right-hand side of
	the \KB equations. The values of these initial correlation functions for a thermal state can
	be determined iteratively in accordance with the underlying 2PI approximation.

%
\section{Numerical Results}\label{ROOE:sec:Numerics}
%

	\begin{figure*}
		\includegraphics[width=0.7\textwidth,keepaspectratio]{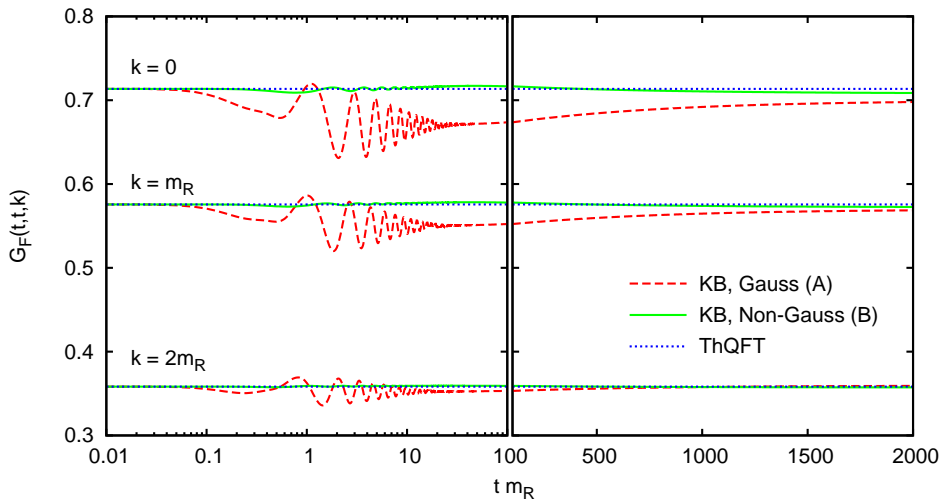}
		\caption{\label{ROOE:fig:GFttThermalInitCorr}\captionstyle
			Time evolution of the equal-time propagator $G_F(t,t,\bm{k})$
			obtained from \KB equations with thermal initial $2$-point correlation
			function (initial state (A), red dashed lines) as well as thermal initial
			$2$- and $4$-point correlation functions (initial state (B), green solid
			lines), for three momentum modes, respectively. The dotted horizontal lines
			show the renormalized thermal propagator $\Gth(0,0,\bm{k})$ which serves as
			initial condition at $t=0$.
		}
	\end{figure*}
	In this section, we support the findings of the previous section by numerical solutions of
	\KB equations. Of course, one cannot implement the complete infinite hierarchy of initial
	$n$-point correlation functions, which would be required for an exact description of thermal
	equilibrium in the framework of \KB equations. However, thermal equilibrium can be approached
	closer and closer as one includes more and more thermal initial correlations. Therefore, we
	consider initial states which are obtained by keeping thermal initial $n$-point correlation
	functions for $n \leq n_{max}$ and setting all higher initial correlations to zero,
	\[
		\alpha_n \left( x_1, \ldots, x_n \right) = \left\{ \begin{array}{ccl}
		\alpha_n^{th} \left( x_1, \ldots, x_n \right) & \mbox{for} & n \leq n_{max} \;, \\
		0                                             & \mbox{for} & n > n_{max} \;.
		\end{array} \right.
	\]
	More precisely, we compare the time evolution obtained for two $Z_2$-symmetric initial states
	(A) and (B) with $n_{max} = 2$ and $n_{max} = 4$, respectively. Both initial states have in
	common that the statistical propagator is initialized with the thermal propagator and that
	all initial correlations $\alpha_n (x_1, \ldots, x_n)$ with $n > 4$ vanish:
	\begin{eqnarray*}
		                              G_F(x^0,y^0,\bm{k})|_{ x^0 = y^0 = 0 } & = & \Gth(-i\tau,0,\bm{k})|_{ \tau \rightarrow 0 } \;, \\
		               \partial_{x^0} G_F(x^0,y^0,\bm{k})|_{ x^0 = y^0 = 0 } & = & 0 \;, \\
		\partial_{x^0} \partial_{y^0} G_F(x^0,y^0,\bm{k})|_{ x^0 = y^0 = 0 } & = & \partial_{\tau}^2 \Gth(-i\tau,0,\bm{k})|_{ \tau\rightarrow 0 } \;, \\
		                             \alpha_n \left(x_1, \ldots, x_n \right) & = & 0 \qquad \mbox{, for } n > 4
	\end{eqnarray*}
	where $\Gth(-i\tau, 0, \bm{k})$ is the complete thermal propagator at temperature
	$T = T_{init}$. The only difference is in the initial 4-point correlation:
	\[
		\alpha_4 \left( x_1, \ldots, x_4 \right) = \left\{ \begin{array}{cl}
		0                                             & \mbox{for (A)} \;, \\
		\alpha_4^{th} \left( x_1, \ldots, x_4 \right) & \mbox{for (B)} \;,
		\end{array} \right.
	\]
	where $\alpha_4^{th} \left( x_1, \ldots, x_4 \right)$ is chosen as shown in
	eq.~(\ref{ROOE:PiThLambdaAlpha}). Accordingly, (A) is a Gaussian initial state and (B)
	is a minimal non-Gaussian initial state, and both states ``are as thermal as possible''
	for the respective classes of initial states.

	In both cases, we employ the 2PI three-loop approximation as discussed in
	Ref.~\cite{Lindner:2005kv} and section~\ref{ROOE:sec:EAaKBE}. We note that
	an equivalent set of equations can be obtained from the 4PI three-loop
	approximation~\cite{Berges:2004pu}. In this case, $\alpha_4$ determines
	the initial value of the 4PI 4-point function~\cite{Garny:PhD}.

	The numerical solutions were obtained on a lattice with $32^3 \times 2000^2$ lattice sites
	and lattice spacings of $a_s m_R = 0.5$ and $a_t m_R = 0.025$. We use $\lambda_R/4!=0.75$
	for the renormalized coupling constant%
	\footnote{%
		We checked for our lattice settings that this value for the
		coupling is far below the `Landau pole' exhibited by the 2PI 3-loop
		approximation~\cite{Berges:2004hn}. Apart from that, we also
		checked that the damping rates for the unequal-time propagators
		increase like $\lambda^2$, which indicates that the above value
		for the coupling lies in the domain of validity of the 2PI 3-loop
		approximation.
	}.
	The bare mass $m_B$ and coupling $\lambda_B$, as well
	as the thermal propagator $\Gth(-i\tau,0,\bm{k})$ are determined by a separate numerics
	program as described in Refs.~\cite{Berges:2004hn,Berges:2005hc}.
	
	\begin{figure*}
		\includegraphics[width=0.7\textwidth,keepaspectratio]{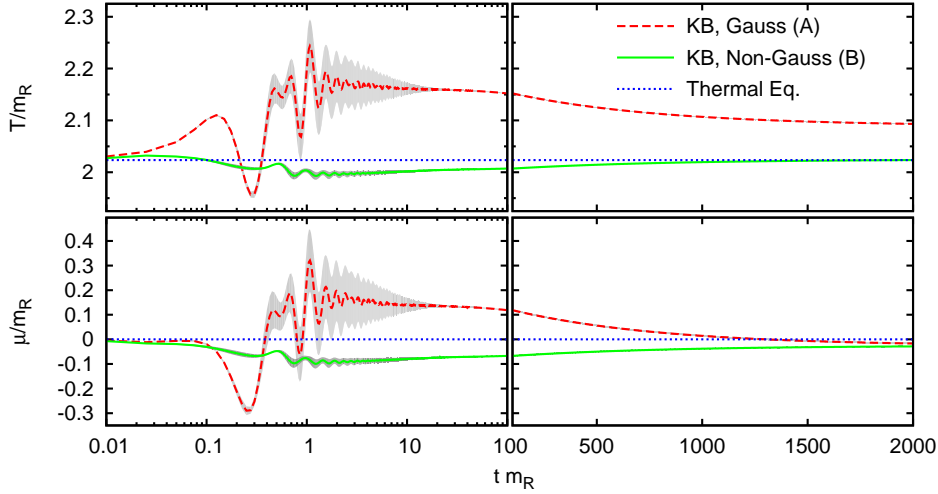}
		\caption{\label{ROOE:fig:BEfit}\captionstyle
			Time evolution of the effective temperature and effective chemical potential obtained
			from \KB equations with thermal initial $2$-point correlation function
			(initial state (A), red dashed lines) as well as thermal initial $2$- and
			$4$-point correlation functions (initial state (B), green solid lines).
			The shaded areas illustrate qualitatively the deviation of the effective
			particle number density $n(t,\bm{k})$ from the Bose-Einstein distribution
			function. They are obtained from the asymptotic standard error of the fit
			(via least-square method) magnified by a factor $10$, for better
			visibility. Nevertheless, the errors become invisibly small at times
			$t m_R\gg 10$.
		}
	\end{figure*}
	%

	\subsection{Time-evolution of the equal-time propagator}
	
	Both initial states (A) and (B) are nonequilibrium initial states, since the thermal initial
	correlations higher than $n_{max}$ are neglected.
	However, among all possible Gaussian initial states, the state (A) has the minimal offset
	from thermal equilibrium. Similarly, the state (B) has the minimal offset from equilibrium
	among all possible initial states that are parameterized by initial $2$- and $4$-point
	correlation functions.
	In the following, we will compare this \emph{minimal offset} for both initial states (A) and (B).
	In exact thermal equilibrium, the equal-time propagator $G_F(t,t,\bm{k})$ is constant.
	Therefore, the time-dependence of the equal-time propagator is a measure for the deviation
	from thermal equilibrium.
	
	Figure~\ref{ROOE:fig:GFttThermalInitCorr} shows the time-evolution of the
	equal-time propagator obtained from the numerical solution of the two sets
	of \KB equations. We find that the deviation from thermal equilibrium is considerable for the
	\KB equations with Gaussian initial state (A). In contrast to this, for the \KB equations with
	the non-Gaussian initial state (B) the equal-time propagator remains very close to the thermal
	propagator at all times.
	We have checked that this qualitative behaviour stays the same when varying the
	coupling strength or the lattice spacings. This shows that the thermal initial $4$-point
	correlation already yields a reasonable approximation of the complete thermal initial state.
	
	This observation can be understood by analyzing the role of the thermal initial
	correlations~(\ref{ROOE:ThermalSelfEnergyCTPSettingSun}) in the \KB
	equations~(\ref{ROOE:KBENonGaussGFGRho}). In fact, there are two distinct
	reasons why the contribution of the thermal initial $n$-point correlations
	are suppressed for $n>4$.
	
	The first reason is that the effective loss of memory is stronger the larger $n$.
	This can be seen as follows: The thermal initial $n$-point correlations enter the \KB
	equations as effective $n$-point vertices. These are connected with classical
	vertices by $n$ lines. Each of these lines yields a propagator
	$G(t,0,\bm{k})$ for which one of the time arguments is evaluated at the initial time
	(see eq.~(\ref{ROOE:AlphaSource})). However, such unequal-time propagators are damped
	exponentially with respect to the time $t$~\cite{Berges:2000ur}. Therefore the contribution of the
	initial $n$-point correlations is also damped exponentially for $t\rightarrow\infty$,
	and the damping is the stronger the larger $n$.
	
	The second reason is that the contributions to the \KB equations~(\ref{ROOE:KBENonGaussGFGRho})
	which arise from thermal initial $n$-point correlations with $n>4$, vanish in the limit
	$t \rightarrow 0$ as has been shown in section~\ref{ROOE:sec:KBEThermalCTP}.
	
	Thus, the influence of thermal initial $n$-point correlations with $n>4$ on the solutions of
	\KB equations is suppressed compared to the initial $4$-point correlation for
	early times ($t\rightarrow 0$) as well as for late times ($t\rightarrow\infty$).
	
	\subsection{Offset between initial and final temperature}
	
	Figure~\ref{ROOE:fig:BEfit} shows the time evolution of the \emph{effective temperature} $T(t)$
	and the \emph{effective chemical potential}%
	\footnote{
		Note that the \emph{effective chemical potential} is introduced here as a fit-parameter in
		order to characterize the equilibration process. Since particle number is not conserved
		for $\lambda\Phi^4$-theory, the \emph{effective chemical potential} has to vanish in thermal
		equilibrium.
	}%
	$\mu(t)$. They are obtained by fitting a Bose-Einstein distribution
	function to the effective particle number density $n(t,\bm{k})$ for all times $t$.
	The effective particle number density can directly be extracted from solutions
	of \KB equations~\cite{Lindner:2005kv}. At the initial time, we have $T(0)=T_{init}$
	and $\mu(0)=0$. For $t\rightarrow\infty$, the solutions approach thermal equilibrium,
	i.e. $T(t)\rightarrow T_{final}$ and $\mu(t)\rightarrow 0$ within the numerical accuracy.
	The time-evolution can be divided into the three phases of
		(i) correlation build-up for $t m_R \lesssim 1$,
		(ii) kinetic equilibration for $1 \lesssim t m_R \lesssim 10$ and
		(iii) chemical equilibration for $t m_R \gtrsim 10$.
	The numerical solutions clearly exhibit a separation between the time-scales of
	kinetic and chemical equilibration~\cite{Berges:2004ce,Lindner:2005kv}.
	
	We see in figure~\ref{ROOE:fig:BEfit} that the deviation from thermal equilibrium is much
	smaller for the non-Gaussian initial state (B) as compared to the Gaussian initial state (A).
	This result supports the observation from the previous subsection.
	
	In addition, figure~\ref{ROOE:fig:BEfit} reveals a qualitative difference between both sets
	of equations. For the Gaussian initial condition (A), there is an offset between the initial
	temperature $T_{init}$ and the final temperature $T_{final}$.
	However, for the non-Gaussian initial condition (B), the initial and final value
	of the temperature agree within the numerical accuracy.
	
	This behaviour can be explained by the composition of the total energy of the system.
	It can be split into kinetic and correlation energy,
	\begin{equation}\label{ROOE:Etotal}
		E_{\it total} = E_{\it kin}(t) + E_{\it corr}(t) \ ,
	\end{equation}
	where
	\begin{eqnarray*}
		\lefteqn{
			E_{\it kin}(t) \ = \ \frac{1}{2} \int\!\!\!\frac{d^3k}{(2\pi)^3}\, \Big[ \partial_{x^0}\partial_{y^0} + \bm{k}^2 + m^2
		}\nonumber\\
		& & {} + \frac{\lambda}{4} \int\!\!\! \frac{d^3q}{(2\pi)^3} G_F(t,t,\bm{q}) \Big] G_F(x^0,y^0,\bm{k})|_{x^0=y^0=t} \ , \nonumber
	\end{eqnarray*}
	\begin{eqnarray*}
		E_{\it corr}(t) & = & {} - \frac{1}{4} \int\!\!\!\frac{d^3k}{(2\pi)^3}\, \bigg[ \Big( \Pi_{\lambda\alpha,\,F}(t,\bm{k}) G_F(0,t,\bm{k}) \nonumber \\
		                &   & {} + \frac{1}{4}\Pi_{\lambda\alpha,\,\rho}(t,\bm{k})G_\rho(0,t,\bm{k})\Big) \nonumber\\
		                &   & {} - \intl_0^t\!\!dz^0\,\Big(\Pi_F(t,z^0,\bm{k})G_\rho(z^0,t,\bm{k})\nonumber\\
		                &   & {} - \Pi_\rho(t,z^0,\bm{k})G_F(z^0,t,\bm{k}) \Big) \bigg] \ .
	\end{eqnarray*}
	The total energy is conserved by the numerical solutions up to numerical errors ($<1\%$).
	Using the thermal initial state, we can also derive an expression for the total energy
	in thermal equilibrium at temperature $T$,
	\[
		E^{\it eq}(T) =  E^{\it eq}_{\it kin}(T) + E^{\it eq}_{\it corr}(T) \ ,
	\]
	where $E^{\it eq}_{\it kin}(T) = E_{\it kin}(t=0)|_{G=\Gth}$ and
	\[
		E^{\it eq}_{\it corr}(T) = - \frac{1}{4}\int\!\!\!\frac{d^3k}{(2\pi)^3}\,
		\Pi_{{\it th},\,\lambda\alpha,\,F}(t,\bm{k})\Gth(0,t,\bm{k})|_{t=0} \ .
	\]
	
	For all possible initial states, the final temperature can then be
	determined by the requirement
	\[
		E_{\it total} = E^{\it eq}(T_{final}) \ .
	\]
	Both initial states considered here feature thermal $n$-point correlations at
	temperature $T_{init}$ for $n\leq n_{max}$. Therefore, we have
	\[
		E_{\it total} = E^{\it eq}(T_{init}) - \Delta E_{(n>n_{max})}(T_{init}) \ ,
	\]
	where $\Delta E_{(n>n_{max})}(T)$ denotes the contribution to the thermal energy
	that comes from $n$-point correlations with $n>n_{max}$.
	For initial state (A), we find
	\[
		\Delta E_{(n>2)}(T) = E^{\it eq}_{\it corr}(T)\ .
	\]
	Since the thermal correlation energy has a  non-zero value, we find that
	\[
		T_{final} \not= T_{init} \qquad \mbox{for initial state (A).}
	\]
	This is in accordance with the results of Ref.~\cite{Kohler:1995zz} for the
	non-relativistic case.
	In general, one would expect that this is also true for initial state (B).
	However, using eq.~(\ref{ROOE:PiThAtInitialTime}), we obtain
	\begin{eqnarray*}
		\lefteqn{
			E^{\it eq}_{\it corr}(T) =
			\frac{\lambda}{4!} \intl_\C d^4x_{1234} \; \Gth(x,x_1)\Gth(x,x_2)
		}\\
		& & {} \times\Gth(x,x_3)\,i\alpha^{\it th}_{4}(x_1,x_2,x_3,x_4)
			\Gth(x_4,x)|_{x=0}\ .
	\end{eqnarray*}
	Thus, only the initial $4$-point correlation contributes to the thermal correlation
	energy. Therefore we have $\Delta E_{(n>4)}(T) = 0$, which means that the initial and
	final temperatures have to agree,
	\[
		T_{final} = T_{init} \qquad \mbox{for initial state (B).}
	\]
	This property is quite remarkable. It means that the total energies for the initial state (B)
	and for the complete thermal initial state are identical. Thus, the thermal initial
	$n$-point correlations $\alpha_n^{\it th}$ with $n>4$ do not contribute to the total
	energy of the initial state. Instead, the thermal initial $4$-point correlation
	$\alpha_4^{\it th}$ already captures the complete thermal correlation energy at
	the initial time in setting-sun approximation.

\begin{figure}
		\feyn{0mm}{1.0}{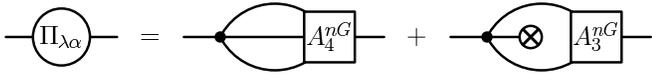}
		\caption{\label{ROOE:fig:PiLambdaAlpha}\captionstyle%
			Contribution $\Pi_{\lambda\alpha}(x,y)$ to the self-energy $\Pi(x,y)$ where the
			left line is connected to a classical vertex, and the right line to an effective non-local vertex.
		}
	\end{figure}

	Moreover, this result can be generalized to arbitrary 2PI approximations. In general, the
	correlation energy at the initial time is given by
	\[
		E_{\it corr}(t=0) = - \frac{1}{4}\int\!\!\!\frac{d^3k}{(2\pi)^3}\,
		\Pi_{\lambda\alpha,\,F}(t,\bm{k}) G_F(0,t,\bm{k})|_{t=0} \ .
	\]
	The most general structure of the non-Gaussian self-energy
	$\Pi_{\lambda\alpha}(x,y)$ is shown in figure~\ref{ROOE:fig:PiLambdaAlpha}.
	The Kernels $A_4^{nG}$ and $A_3^{nG}$ may in general contain classical as well
	as non-local effective vertices. In setting-sun
	approximation~(\ref{ROOE:Gamma2ThreeLoopWithInitial4PointSource}), they are given by
	$A_4^{nG}=\alpha_4$ and $A_3^{nG}=0$.
	In the limit $t\equiv x^0 \rightarrow 0$, all contributions containing classical vertices
	vanish due to the memory integrals accompanying these vertices. Thus, at the initial time,
	only those parts of $A_4^{nG}$ and $A_3^{nG}$ do contribute to the correlation energy
	that can be rewritten in the form of an initial $4$- and $3$-point vertex, respectively.

	This property is characteristic for the $\Phi^4$-interaction. If the Lagrangian would contain a
	(non-renormalizable) $\Phi^6$-interaction, then also initial $5$- and $6$-point correlations
	would explicitly contribute to the energy density of the thermal initial state.

%
\section{Conclusions and Outlook}
%

	In this work, we derive \KB equations for non-Gaussian correlated initial states describing
	vacuum and thermal equilibrium. We provide suitable techniques to establish approximations to
	the exact thermal initial correlation functions that match the approximation scheme underlying
	the \KB equations. These techniques are applicable for arbitrary truncations of the 2PI
	effective action. Examples are given for the 2PI three-loop approximation. 
	
	Finally, we discuss numerical solutions of \KB equations for a real scalar $\Phi^4$ quantum
	field theory, which take the thermal initial $4$-point correlation as the leading non-Gaussian
	correction into account. These solutions are compared to solutions obtained for Gaussian initial
	states.
	For the latter, the initial state has no correlation energy by definition.
	Therefore, even if one initializes the two-point function with the thermal propagator for a
	certain temperature, the system equilibrates at a different temperature~\cite{Kohler:1995zz}.
	We show numerically and analytically that this feature of the Gaussian initial state is
	remedied completely already by taking the thermal initial $4$-point correlation into account.
	The reason is that higher correlations of the initial state do not contribute to the total
	energy at the initial time. Thus, we find that including an initial $4$-point correlation
	function yields a significantly improved approximation to the complete thermal initial state
	as compared to Gaussian initial states.
	
	The techniques developed in this work provide a framework for investigating the
	renormalization of \KB equations. It is known that the 2PI effective action can be
	renormalized in thermal equilibrium~\cite{Berges:2005hc,Berges:2004hn}. Accordingly, it is
	possible to derive renormalized \KB equations for the thermal initial state using the
	techniques introduced above. These provide a well-defined \emph{expansion point} for
	nonequilibrium initial states. For example, it is possible to parameterize the initial
	$n$-point correlation functions in the form
	$
		\alpha_n = \alpha_n^{\it th} + \Delta\alpha_n\ .
	$
	In the limit $\Delta\alpha_n\rightarrow 0$, the thermal initial state is recovered and thus
	the \KB equations are formally finite. Then it is possible to investigate which conditions
	the deviations $\Delta\alpha_n$ have to fulfill such that no new divergences are introduced.
	This is left to future work.

\begin{acknowledgments}
	The numerical solutions were obtained on the HLRB II (SGI Altix 4700) at
	the Leibniz Supercomputing Centre. We thank J\"urgen Berges, Szabolcs Borsanyi,
	Guy Moore, Emil Mottola and Urko Reinosa for helpful comments and discussions.
\end{acknowledgments}

\begin{appendix}
%
\section{Perturbative Calculation of Thermal Density Matrix Element}\label{QFOOE:sec:ThermalInitialState}
%
	
	In order to describe the equilibrium limit within non\-equilibrium quantum field theory, it is
	important to calculate the thermal density matrix element
	$\left\langle\varphi_+\left|\rho_{th}\right|\varphi_-\right\rangle$ evaluated with
	respect to two eigenstates
	$\Phi(0,\bm{x})|\varphi_\pm\rangle = \varphi_\pm(\bm{x})|\varphi_\pm\rangle$ of the field
	operator $\Phi(t,\bm{x})$ at time $t=t_{init}\equiv 0$. In the following, the perturbative
	expansion of the thermal density matrix element is discussed based on
	Ref.~\cite{Calzetta:1986cq}. Therefore, the action~(\ref{ROOE:Action}) is formulated on the
	imaginary time path, and split into the free part
	$S_0[\phi_0]=\int_\I d^4x\left((\partial\phi_0)^2-m^2\phi_0^2\right)/2$ and the interaction
	part $S_{int}[\phi]=-\int_\I d^4x\frac{\lambda}{4!}\phi(x)^4$.
	For the free thermal density matrix $\rho_0 = \frac{1}{Z_0}\,\exp\left(-\beta H_0\right)$
	containing the free Hamiltonian $H_0$, which is quadratic in the field, the matrix element
	can be calculated by a path integral that is analogous to
	eq.~(\ref{QFOOE:ThermalDensityMatrixElement}). The result is~\cite{Calzetta:1986cq}
	\[
		\left\langle \varphi_+ \left| \rho_0 \right| \varphi_- \right\rangle
		= \mathcal{N}_0 \exp\left[ iS_0[\phi_0]\rule{0mm}{4mm} \right]\,,
	\]
	where $\mathcal{N}_0$ is a normalization factor, which is independent of $\varphi_\pm$, and
	$\phi_0(x)$ is the solution of the free equation of motion
	$\delta S_0/\delta\phi=(-\Box-m^2)\phi_0=0$ on $\I$ subject to the boundary conditions
	\[
		\phi_0(0,\bm{x}) = \varphi_-(\bm{x}) \quad \mbox{and} \quad \phi_0(-i\beta,\bm{x})=\varphi_+(\bm{x})\,.
	\]
	The solution is uniquely determined, and, in spatial momentum space, given by
	\[
		\phi_0(-i\tau,\bm{k}) = \frac{\sinh(\wk\tau)}{\sinh(\wk\beta)}\varphi_+(\bm{k})
		+ \frac{\sinh(\wk(\beta-\tau))}{\sinh(\wk\beta)}\varphi_-(\bm{k})\,,
	\]
	where $\wk^2=m^2+\bm{k}^2$. 
	
	The full thermal initial correlations can be obtained by perturbing the full Hamiltonian $H$
	around $H_0$,
	\[
		\left\langle \varphi_+ \left| \rho_{th} \right| \varphi_- \right\rangle
		= \mathcal{N}\exp\left[i\left(S_0[\phi_0]+F_{int}[\phi_0]\right)\rule{0mm}{4mm}\right]\,,
	\]
	where $\mathcal{N}$ is a normalization factor and $iS_0[\phi_0]$ is the free contribution.
	$iF_{int}[\phi_0]$ is the sum of all connected Feynman diagrams with vertices given by the
	derivatives of $S_{int}[\phi]$ evaluated for $\phi(x)=0$,
	\[
		\frac{i\delta^4S_{int}}{\delta\phi^4}=-i\lambda\delta_\I(x_1-x_2)\delta_\I(x_1-x_3)\delta_\I(x_1-x_4)
		\ = \ \feyn{-2.5mm}{0.075}{VertexI} \;.
	\]
	According to the Feynman rules given in section \ref{ROOE:sec:PerturbativeThermInitCorr}
	the empty circle reminds us that the corresponding integration runs over the imaginary-time
	contour $\I$.
	The boundary conditions of the path integral~(\ref{QFOOE:ThermalDensityMatrixElement}) are
	formally taken into account by the ``field expectation value''
	\[
		\phi_0(-i\tau,\bm{k}) \ = \quad \feyn{-1mm}{0.2}{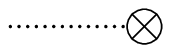}\ ,
	\]
	along the imaginary contour $\I$, as well as the propagator
	\begin{eqnarray}
		\lefteqn{D_0 (-i\tau, -i\tau', \bm{k}) = \feyn{0.5mm}{0.25}{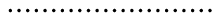}} \label{QFOOE:D0Prop} \\
		& = & \frac{1}{\wk\sinh(\wk\beta)} \Big( \sinh(\wk\tau) \sinh(\wk(\beta - \tau')) \Theta(\tau' - \tau) \nonumber \\
		&   & {} + \sinh(\wk\tau') \sinh(\wk(\beta - \tau)) \Theta(\tau - \tau') \Big) \;, \nonumber
	\end{eqnarray}
	which is the Greens function for solutions of the free equation of motion that vanish at the
	boundaries $\tau=0,\beta$. $D_0$ is denoted by the dotted line. To first order in
	$\lambda$, $iF_{int}[\phi_0]$ is given by
	\begin{eqnarray*}
		\lefteqn{\left. iF_{int}[\phi_0]\right|_{\mathcal{O}(\lambda)} = \feyn{-2.0mm}{0.21}{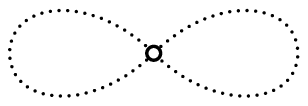}
		                                                               + \feyn{-6mm}{0.19}{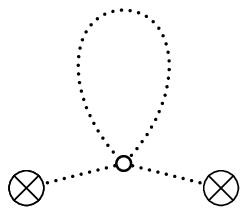}
		                                                               + \feyn{-6mm}{0.15}{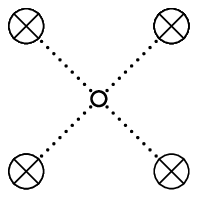} } \\
		& = & \frac{-i\lambda}{4!} \intl_{\I} d^4 x \left[ 3D_0^2 (x,x) + 6\phi_0^2 (x) D_0(x,x) + \phi_0^4 (x) \right] \;.
	\end{eqnarray*}
	The field-independent diagrams, like the first one above, can be absorbed into the
	normalization $\mathcal{N}$. The perturbative expansions of the thermal initial
	correlations $\alpha_n^{\it th}$ are obtained by the $n$-th functional derivative of
	$F[\varphi_+,\varphi_-] \equiv S_0[\phi_0] + F_{int}[\phi_0]$ with respect to the field,
	\[ i\alpha_n^{th, \epsilon_1, \ldots, \epsilon_n} (\bm{x_1}, \ldots, \bm{x_n})
		= \left. \frac{\delta iF[\varphi_+,\varphi_-]}{\delta\varphi_{\epsilon_1} (\bm{x_1}) \ldots \delta\varphi_{\epsilon_n} (\bm{x_n})}\right|_{\varphi=0} \;,
	\]
	to which all diagrams with $n$ insertions of $\phi_0$ contribute. Here, the decomposition
	from eq.~(\ref{ROOE:AlphaSource}) has been used. Thus, the initial correlations obtained
	in this way are indeed supported only at the initial time, as required.
	Formally, the functional derivative corresponds to replacing the field insertions by
	(distinguishable) external lines in the diagrammatic expansion of $iF_{int}[\phi_0]$
	according to
	\begin{eqnarray*}
		\phi_0(-i\tau,\bm{k})\quad &\mapsto& \Delta_0(-i\tau,x^0,\bm{k})\\
		\feyn{-2mm}{0.2}{DottedPhi0}
		\quad &\mapsto& \ \feyn{-2mm}{0.2}{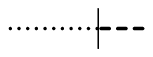}\nonumber\ ,
	\end{eqnarray*}
	where $\Delta_0(-i\tau,x^0,\bm{k})$ is defined in eq.~(\ref{FreeDelta}).
	For example, the leading contribution to the thermal initial four-point correlation
	function obtained from the fourth derivative of $iF_{int}[\phi_0]$ is given by
	(see eq.~(\ref{ROOE:FourierTrfOfDelta}))
	\begin{eqnarray}\label{QFOOE:ThermalAlpha4}
		\lefteqn{i\alpha_{4,\, 0L}^{\it th}(x_1,x_2,x_3,x_4) = \qquad \feyn{-5mm}{0.14}{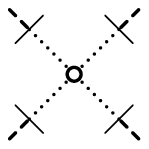}}\\
		& = & -i\lambda \intl_{\I} d^4 v \; \Delta_0(v,x_1) \Delta_0(v,x_2) \Delta_0(v,x_3) \Delta_0(v,x_4) \;. \nonumber
	\end{eqnarray}
	Switching again to momentum space, an explicit expression for the leading contribution to the
	perturbative thermal initial four-point correlation function is obtained,
	\begin{eqnarray*}
		i\alpha_{4, 0L}^{th, \epsilon_1, \epsilon_2, \epsilon_3, \epsilon_4} (\bm{k_1}, \bm{k_2}, \bm{k_3}, \bm{k_4})
		& = & - \lambda \intl_0^\beta d \tau \; \Delta_0^{\epsilon_1}(-i\tau,\bm{k_1}) \\
		\lefteqn{
			\hspace*{-25mm}\times\Delta_0^{\epsilon_2}(-i\tau,\bm{k_2})\Delta_0^{\epsilon_3}(-i\tau,\bm{k_3})\Delta_0^{\epsilon_4}(-i\tau,\bm{k_4}) \;,
		}\nonumber
	\end{eqnarray*}
	where
	\begin{eqnarray*}
		\Delta_0^+ (-i \tau, \bm{k}) & = & \frac{\sinh(\wk\tau)}{\sinh(\wk\beta)} \;, \\
		\Delta_0^- (-i \tau, \bm{k}) & = & \frac{\sinh(\wk(\beta - \tau))}{\sinh(\wk\beta)} \;.
	\end{eqnarray*}
	The integral over the imaginary time can be performed analytically.
	In the zero-temperature limit ($\beta\rightarrow\infty$), one obtains
 	\begin{eqnarray}
		\lefteqn{i\alpha_{4, 0L}^{th, \epsilon_1, \epsilon_2, \epsilon_3, \epsilon_4} (\bm{k_1}, \bm{k_2}, \bm{k_3}, \bm{k_4})} \label{QFOOE:Alpha4Perturbative} \\
		& = & \left\{ \begin{array}{cl}
		      \frac{-\lambda}{\wks{1}+\wks{2}+\wks{3}+\wks{4}} & \ \mbox{for}\ \epsilon_1=\epsilon_2=\epsilon_3=\epsilon_4= \pm \;, \\
		                                                     0 & \ \mbox{else} \;.
		      \end{array} \right. \nonumber
	\end{eqnarray}
	Altogether, a diagrammatic expansion of the matrix element of the thermal density matrix in
	terms of perturbative Feynman diagrams has been developed as suggested in
	Ref.~\cite{Calzetta:1986cq}. This allows to explicitly calculate thermal correlation functions
	order by order in the quartic coupling constant. The lowest-order perturbative
	result~(\ref{QFOOE:ThermalAlpha4}) may be compared to the nonperturbative 2PI
	result~(\ref{ROOE:ThermalAlpha4}).

%
\section{Derivation of the connection in the 2PI case}\label{QFOOE:sec:FullConnection}
%
	
	In this appendix, we generalize eq.~(\ref{GthCDelta}) to the 2PI formalism. Before we consider
	the complete thermal propagator, we find it helpful to take an intermediate step by first
	considering a \emph{mixed} thermal propagator which is identical to the complete thermal
	propagator on the imaginary branch of the thermal time contour, and which obeys the free equation
	of motion on the real branches.
	
	\subsection{Mixed thermal propagator}
	
	It is straightforward to define projectors on the parts $\C$ and $\I$ of the thermal time
	contour,
	\[ \EI(x^0) = \left\{\begin{array}{l@{\quad\mbox{if}\quad}l}
				0 & x^0 \in \C\\
				1 & x^0 \in \I
			\end{array}\right. \]
	and
	\[ \EC(x^0) = \left\{\begin{array}{l@{\quad\mbox{if}\quad}l}
				1 & x^0 \in \C\\
				0 & x^0 \in \I
			\end{array}\right. \, .\]
	Of course, they fulfill the relation
	\[
		\EI(x^0) + \EC(x^0) = 1 \qquad \mbox{for all } x^0\in\CpI\ .
	\]
	The mixed thermal propagator is defined by the following equation of motion,
	\begin{eqnarray}
		\GthM^{-1}(x,y) & = & i(\Box_x+m^2)\delta_{\CpI}(x-y) \nonumber \\
				&   & {} - \EI(x^0)\EI(y^0)\Pith(x,y) \, , \label{ROOE:GthMixed}
	\end{eqnarray}
	where $x^0, y^0\in\CpI$. Here $\Pith(x,y)$ is the complete thermal self-energy. The mixed
	propagator can be decomposed into statistical and spectral components,
	\[
		\GthM(x,y) = \GthMF(x,y) - \frac{i}{2} \sgn_{\CpI}(x^0-y^0) \GthMR(x,y) \,.
	\]
	The equation of motion for the mixed propagator can equivalently be written as
	\begin{eqnarray*}
		\lefteqn{\left( \Box_x + m^2 \right) \GthM (x,y) = -i \delta_{\CpI} (x - y)} \\
                & & {} - i \EI(x^0) \intl_{\I} d^4 z \; \PithM(x,z) \GthM(z,y) \;.
	\end{eqnarray*}
	Each of the two time arguments of the propagator can either be real or imaginary, which
	yields four combinations $\GthM^{\C\C}$, $\GthM^{\C\I}$, $\GthM^{\I\C}$, $\GthM^{\I\I}$. The
	mixed propagator evaluated with two imaginary time arguments is identical to the complete thermal
	2PI propagator,
	\begin{equation}\label{ROOE:GthMII}
		\GthM^{\I\I}(x,y) = \Gth^{\I\I}(x,y) \qquad \mbox{for } x^0, y^0\in\I,
	\end{equation}
	whereas the mixed propagators evaluated with one or two real time arguments, $\GthM^{\C\I}(x,y)$ and
	$\GthM^{\C\C}(x,y)$, fulfill the equation of motion of the free propagator,
	\begin{eqnarray}\label{ROOE:EoMGthMixedCIandCC}
		\left(\Box_x + m^2\right)\GthM^{\C\I}(x,y)  & = & 0 \,, \\
		\left(\Box_x + m^2\right)\GthMF^{\C\C}(x,y) & = & \left(\Box_x + m^2\right)\GthMR^{\C\C}(x,y) = 0 \,. \nonumber
	\end{eqnarray}
	At the initial time $x^0=y^0=0$, the propagators on all branches of the thermal time path
	agree. Using eq.~(\ref{ROOE:GthMII}), one obtains
	\[
		\GthM^{\mathcal{P}_1\mathcal{P}_2}(x,y)|_{x^0=y^0=0} = \Gth(x,y)|_{x^0=y^0=0} \,,
	\]
	for $\mathcal{P}_i\in\{\C,\I\}$. Thus, the initial value of the mixed propagator at
	$x^0=y^0=0$ is given by the complete thermal propagator.
	
	For the mixed propagator with one imaginary and one real time, the equation of motion,
	transformed to spatial momentum space, reads
	\begin{eqnarray*}
		\lefteqn{\left(-\partial_{\tau}^2 + \bm{k}^2 + m^2\right)\GthM^{\I\C}(-i\tau, y^0, \bm{k})} \\
		& = &  - \intl_0^{\beta} d \tau' \; \Pith^{\I\I} (-i\tau, -i\tau', \bm{k}) \GthM^{\I\C} (-i\tau', y^0, \bm{k}) \;.
	\end{eqnarray*}
	Next, a Fourier transformation with respect to the imaginary time is performed, using in
	particular
	\begin{eqnarray*}
		\lefteqn{\intl_0^{\beta} d \tau \; \exp \left( -i \omega_n \tau \right) \partial_{\tau}^2 \GthM^{\I\C} (-i\tau, y^0, \bm{k})} \\
		& = & - \omega_n^2 \GthM^{\I\C} (\omega_n, y^0, \bm{k}) \\
		&   & {} + \disc(i \omega_n \GthM^{\I\C} + \partial_{\tau} \GthM^{\I\C}) (y^0, \bm{k}) \;,
	\end{eqnarray*}
	where $\omega_n=2\pi\beta n$ is a Matsubara frequency. It is important to take the
	contribution from boundary terms into account,
	\begin{eqnarray*}
		\lefteqn{\disc(i\omega_n\,\GthM^{\I\C}+\partial_\tau\GthM^{\I\C})(y^0,\bm{k})} \\
		& = & \left[(i\omega_n\,\GthM^{\I\C}+\partial_\tau\GthM^{\I\C})(-i\tau,y^0,\bm{k})\right]_{\tau=0}^{\tau=\beta} \;.
	\end{eqnarray*}
	Thus, the Fourier transformed equation for the mixed propagator reads
	\begin{eqnarray}\label{ROOE:GthMixedIC}
		\lefteqn{\left(\omega_n^2 + \bm{k}^2 + m^2\right)\GthM^{\I\C}(\omega_n,y^0,\bm{k})} \\
		& = & {} - \Pith^{\I\I}(\omega_n,\bm{k})\GthM^{\I\C}(\omega_n,y^0,\bm{k}) \nonumber\\
		&   & {} + \disc(i\omega_n\,\GthM^{\I\C}+\partial_\tau\GthM^{\I\C})(y^0,\bm{k}) \;. \nonumber
	\end{eqnarray}
	The boundary terms have to fulfill the equation of motion
	\begin{eqnarray*}
		\left(\partial_{y^0}^2 + \bm{k}^2 + m^2\right)\disc(\GthM^{\I\C})(y^0,\bm{k})              & = & 0 \;, \\
		\left(\partial_{y^0}^2 + \bm{k}^2 + m^2\right)\disc(\partial_\tau\GthM^{\I\C})(y^0,\bm{k}) & = & 0 \;,
	\end{eqnarray*}
	which follows using $\GthM^{\I\C}(\omega_n,y^0,\bm{k})=\GthM^{\C\I}(y^0,\omega_n,\bm{k})$ and
	the equation of motion~(\ref{ROOE:EoMGthMixedCIandCC}) for $\GthM^{\C\I}$. Furthermore, the
	initial conditions at $y^0=0$ are fixed by the periodicity relation of the thermal propagator
	as well as the equal-time commutation relations,
	\begin{eqnarray*}
		\disc(\GthM^{\I\C})(0,\bm{k})                              & = & 0 \;, \\
		\partial_{y^0}\,\disc(\GthM^{\I\C})(0,\bm{k})              & = & i \;, \\
		\disc(\partial_\tau\GthM^{\I\C})(0,\bm{k})                 & = & 1 \;, \\
		\partial_{y^0}\,\disc(\partial_\tau\GthM^{\I\C})(0,\bm{k}) & = & 0 \;.
	\end{eqnarray*}
	The statistical and spectral components $\GthMF^{\C\C}(0,y^0,\bm{k})$ and
	$\GthMR^{\C\C}(0,y^0,\bm{k})$ of the mixed propagator are two linearly independent solutions
	of the free equation of motion. Since it is a second order differential equation, any solution
	can be expressed as a linear combination, especially
	\begin{eqnarray*}
		\disc(\GthM^{\I\C})(y^0,\bm{k})             & = & -i\GthMR^{\C\C}(0,y^0,\bm{k}) \;, \\
		\disc(\partial_\tau\GthM^{\I\C})(y^0,\bm{k})& = & \frac{\GthMF^{\C\C}(0,y^0,\bm{k})}{\Gth(0,0,\bm{k})} \;. \nonumber
	\end{eqnarray*}
	Inserting this result into eq.~(\ref{ROOE:GthMixedIC}) and using the Fourier-transformed
	Schwinger-Dyson equation~(\ref{ROOE:GthFull}) for the complete thermal propagator yields
	\begin{eqnarray*}
		\GthM^{\I\C} (\omega_n, y^0, \bm{k}) & = & \left( \frac{\Gth^{\I\I} (\omega_n, \bm{k})}{\Gth (0, 0, \bm{k})} \right) \GthMF^{\C\C} (0, y^0, \bm{k}) \\
		                                     &   & {} - \left( i \omega_n \Gth^{\I\I} (\omega_n, \bm{k}) \right) \GthMR^{\C\C} (0, y^0, \bm{k}) \;.
	\end{eqnarray*}
	Finally, the upper relation can be rewritten in the form
	\begin{equation}\label{ROOE:GthMDelta}
		\GthM^{\I\C}(\omega_n,y^0,\bm{k}) = \intl_{\C} d z^0 \; \Delta_m (\omega_n, z^0, \bm{k}) \GthM^{\C\C} (z^0, y^0, \bm{k}) \;,
	\end{equation}
	where a \emph{mixed connection} has been introduced,
	\begin{eqnarray}\label{ROOE:DeltaM}
		\lefteqn{
			\Delta_m(\omega_n,z^0,\bm{k})
			= \Delta_m^s(\omega_n,\bm{k})\delta_s(z^0) + \Delta_m^a(\omega_n,\bm{k})\delta_a(z^0)
		}\nonumber\\
		& = & \left( \frac{\Gth^{\I\I}(\omega_n,\bm{k})}{\Gth(0,0,\bm{k})} \right)\delta_s(z^0) + \left( \rule{0mm}{6mm}2i\omega_n\Gth^{\I\I}(\omega_n,\bm{k}) \right)\delta_a(z^0) \nonumber \\
		& = & \qquad \feyn{-1.5mm}{0.2}{DeltaM}\ .
	\end{eqnarray}
	Furthermore, the transposed connection is defined as
	$\Delta_m^T(z^0,\omega_n,\bm{k})=\Delta_m(\omega_n,z^0,\bm{k})$. Eq.~(\ref{ROOE:GthMDelta})
	for the mixed propagator is the generalization of eq.~(\ref{GthCDelta}) for the free propagator.
	Thus, the mixed propagator evaluated with one real and one imaginary time can be written as
	the convolution of the mixed connection, which involves the complete 2PI propagator, and the
	real-real mixed propagator, which obeys the free equation of motion.

	\subsection{Complete thermal propagator}
	
	Using the equation of motion~(\ref{ROOE:GthMixed}) of the mixed propagator, the
	self-consistent equation of motion~(\ref{ROOE:GthFull}) of the complete propagator can be rewritten as
	\[
		\Gth^{-1}(x,y) = \GthM^{-1}(x,y) - \left[1-\EI(x^0)\EI(y^0)\right]\Pith(x,y)\,,
	\]
	for $x^0, y^0\in\CpI$.
	By convolving this equation with $\Gth$ from the left and with $\GthM$ from the right, the
	integral form of the Schwinger-Dyson equation is obtained:
	\begin{eqnarray}
		\lefteqn{\Gth (x, y) = \GthM(x,y) + \intl_{\CpI} d^4 u \intl_{\CpI} d^4 v \; \Gth (x, u)} \nonumber \\
		&   & {} \times \left[ 1 - \EI(u^0) \EI(v^0) \right] \Pith(u,v) \GthM(v,y) \;. \qquad \label{ROOE:SchwingerDysonGthGthM}
	\end{eqnarray}
	Evaluating it for $x^0\in\C$ and $y^0\in\I$, and performing a Fourier transformation with
	respect to the relative spatial coordinate $\bm{x}-\bm{y}$ as well as the imaginary time
	$y^0$ gives
	\begin{eqnarray*}
		\lefteqn{\Gth^{\C\I} (x^0, \omega_n, \bm{k}) = \GthM^{\C\I} (x^0, \omega_n, \bm{k})} \\
		& & {} + \intl_{\CpI} du^0 \intl_{\C} dv^0 \; \Gth(x^0, u^0, \bm{k}) \\
		& & \qquad {} \times \Pith(u^0, v^0, \bm{k}) \GthM^{\C\I}(v^0, \omega_n, \bm{k}) \\
		& & {} - i \intl_{\C} du^0 \; \Gth^{\C\C} (x^0, u^0, \bm{k}) \Pith (u^0, \omega_n, \bm{k}) \GthM^{\I\I} (\omega_n, \bm{k}) \;.
	\end{eqnarray*}
	Next, $\GthM^{\C\I}(x^0,\omega_n,\bm{k})$ and $\GthM^{\C\I}(v^0,\omega_n,\bm{k})$ are replaced
	using eq.~(\ref{ROOE:GthMDelta}) with interchanged arguments. Furthermore, it is used that
	$\GthM^{\I\I}(\omega_n,\bm{k})=\Gth^{\I\I}(\omega_n,\bm{k})$ (see eq.~(\ref{ROOE:GthMII})):
	\begin{widetext}
	\begin{eqnarray*}
		\lefteqn{\Gth^{\C\I} (x^0, \omega_n, \bm{k}) = \intl_{\C} dz^0 \; \GthM^{\C\C} (x^0, z^0, \bm{k}) \Delta_m^T (z^0, \omega_n, \bm{k})} \\
                &   & {} + \intl_{\C} dz^0 \intl_{\CpI} du^0 \intl_{\C} dv^0 \; \Gth (x^0, u^0, \bm{k}) \Pith (u^0, v^0, \bm{k})
		      \GthM^{\C\C} (v^0, z^0, \bm{k}) \Delta_m^T (z^0, \omega_n, \bm{k}) \\
		&   & {} - i\intl_{\C} du^0 \; \Gth^{\C\C} (x^0, u^0, \bm{k}) \Pith(u^0, \omega_n, \bm{k}) \Gth^{\I\I} (\omega_n, \bm{k}) \\
		& = & \intl_{\C} dz^0 \left[ \Gth^{\C\C} (x^0, z^0, \bm{k}) - \intl_{\C} du^0 \intl_{\I} dv^0 \; \bigg( \Gth^{\C\C} (x^0, u^0, \bm{k}) \Pith (u^0, v^0, \bm{k})
		      \GthM^{\I\C} (v^0, z^0, \bm{k}) \bigg) \right] \Delta_m^T (z^0, \omega_n, \bm{k}) \\
		&   & {} - i \intl_{\C} du^0 \; \Gth^{\C\C} (x^0, u^0, \bm{k}) \Pith (u^0, \omega_n, \bm{k}) \Gth^{\I\I} (\omega_n, \bm{k}) \\
		& = & \intl_{\C} dz^0 \; \Gth^{\C\C} (x^0, z^0, \bm{k}) \left\{ \rule{0mm}{7mm} \Delta_m^T (z^0, \omega_n, \bm{k})
		      - i \Pith (z^0, \omega_n, \bm{k}) \Gth^{\I\I} (\omega_n, \bm{k}) \right. \\
		&   & \left. \rule{0mm}{7mm} - \intl_{\C} du^0 \intl_{\I} dv^0 \; \Pith (z^0, v^0, \bm{k}) \GthM^{\I\C} (v^0, u^0, \bm{k}) \Delta_m^T (u^0, \omega_n, \bm{k}) \right\} \;.
	\end{eqnarray*}
	\end{widetext}
	In the second step, the Schwinger-Dyson equation~(\ref{ROOE:SchwingerDysonGthGthM}) evaluated
	for $x^0,z^0\in\C$ was used again. In the third step the complete real-real propagator was
	factored out by interchanging the integration variables $u^0\leftrightarrow z^0$ in the
	second and third term. The last line can be simplified by Fourier transforming with respect to
	the imaginary time $v^0$, and performing the integral over $\C$ using eq.~(\ref{ROOE:DeltaM}):
	\begin{eqnarray*}
		\lefteqn{\!\!\!\! \intl_{\C} \!\! du^0 \!\! \intl_{\I} \!\! dv^0 \; \Pith (z^0, v^0, \bm{k}) \GthM^{\I\C} (v^0, u^0, \bm{k}) \Delta_m^T (u^0, \omega_n, \bm{k})} \\
		& = & -iT \sum_l \Pith (z^0, \omega_l, \bm{k}) \GthM^{\I\C} (\omega_l, 0, \bm{k}) \Delta_m^s (\omega_n, \bm{k}) \\
		& = & -iT \sum_l \Pith (z^0, \omega_l, \bm{k}) \Gth^{\I\I} (\omega_l, \bm{k}) \frac{\Gth^{\I\I} (\omega_n, \bm{k})}{\Gth (0, 0, \bm{k})} \;.
	\end{eqnarray*}
	Finally, a decomposition of the complete thermal 2PI propagator evaluated with one real time and
	one Matsubara frequency is obtained,
	\[
		\Gth^{\C\I} (x^0, \omega_n, \bm{k}) = \intl_{\C} dz^0 \; \Gth^{\C\C} (x^0, z^0, \bm{k}) \Delta^T (z^0, \omega_n, \bm{k}) \;,
	\]
	where the \emph{complete connection} was introduced,
	\begin{eqnarray}\label{ROOE:DeltaMatsubara}
	       \lefteqn{\Delta^T(z^0,\omega_n,\bm{k}) = \Delta_m^T(z^0,\omega_n,\bm{k})} \\
	  & &  {} - iT\sum_m\,\Pith(z^0,\omega_m,\bm{k})\,D(\omega_m,\omega_n,\bm{k})\,,\nonumber
	\end{eqnarray}
	with $\Delta(\omega_n,z^0,\bm{k})=\Delta^T(z^0,\omega_n,\bm{k})$.
	Compared to the mixed connection, the complete connection contains an additional term, which is
	the convolution of the thermal self-energy, evaluated with one real time and one Matsubara
	frequency, with the propagator $D(\omega_m,\omega_n,\bm{k})$. This propagator is given by
	\begin{eqnarray}
		\lefteqn{D (\omega_n, \omega_m, \bm{k}) = \frac{\delta_{n,m}}{T} \Gth^{\I\I} (\omega_n, \bm{k})} \nonumber \\
		&   & \hspace{-12pt} {} - \intl_{\C} \!\! dw^0 \!\! \intl_{\C} \!\! dz^0 \; \Delta (\omega_n, w^0, \bm{k}) \Gth(w^0, z^0, \bm{k}) \Delta^T (z^0, \omega_m, \bm{k}) \nonumber \\
		& = & \frac{\delta_{n,m}}{T} \Gth^{\I\I} (\omega_n, \bm{k}) - \frac{\Gth^{\I\I} (\omega_n, \bm{k}) \Gth^{\I\I} (\omega_m, \bm{k})}{\Gth (0, 0, \bm{k})} \;. \label{ROOE:DMatsubara}
	\end{eqnarray}
	In the last line
	\[
		\intl_{\C} \!\! dw^0 \!\! \intl_{\C} \!\! dz^0 \; X(\omega_n, w^0, \bm{k}) \Gth(w^0, z^0, \bm{k}) \Pith(z^0, \omega_m, \bm{k}) = 0
	\]
	was used, where $X\in\{\Delta,\Pith\}$. The propagator $D$ has the properties
	\begin{eqnarray*}
		D(\omega_n, \omega_m, \bm{k}) & = & D(\omega_m, \omega_n, \bm{k}) \;, \\
		T \sum_m D(\omega_n, \omega_m, \bm{k}) & = & 0 \;.
	\end{eqnarray*}
	From the last property it can be inferred that only the non-local part of the thermal
	self-energy $\Pith(z^0,\omega_m,\bm{k})=\Pith^{\it loc}+\Pith^{\it nl}(z^0,\omega_m,\bm{k})$
	contributes in eq.~(\ref{ROOE:DeltaMatsubara}), since the local part is independent of the
	Matsubara frequency.
	
	By applying an inverse Fourier transformation with respect to imaginary time, using in
	particular
	\[ D(-i\tau, -i\tau', \bm{k}) = T^2 \sum_{n, m} e^{i \omega_n \tau - i \omega_m \tau'} D (\omega_n, \omega_m, \bm{k}) \;, \]
	the complete thermal 2PI propagator with one imaginary and one real time can be decomposed as
	\begin{eqnarray}
		\Gth^{\C\I}(x^0,-i\tau,\bm{k})        & = & \!\! \intl_{\C} \!\! dz^0 \; \Gth^{\C\C} (x^0, z^0, \bm{k}) \Delta^T (z^0, -i\tau, \bm{k}) \;, \nonumber \\
		\feyn{0mm}{0.175}{PropagatorCI} \quad & = & \qquad \feynRot{+3mm}{0.22}{180}{PropagatorDelta} \ ,\nonumber
	\end{eqnarray}
	and
	\begin{eqnarray}\label{ROOE:GthDelta}
		\Gth^{\I\C}(-i\tau,y^0,\bm{k})        & = & \!\! \intl_{\C} \!\! dz^0 \; \Delta (-i\tau, z^0, \bm{k}) \Gth^{\C\C} (z^0, y^0, \bm{k}) \;, \nonumber \\
		\feyn{0mm}{0.175}{PropagatorIC} \quad & = & \qquad \feyn{-1mm}{0.22}{PropagatorDelta} \ ,
	\end{eqnarray}
	where the complete connection is given by
	\begin{eqnarray}
		\lefteqn{\Delta(-i\tau, z^0, \bm{k}) = \Delta_m(-i\tau, z^0, \bm{k})} \nonumber \\
		&        & {} + \intl_{\I} dv^0 \; D(-i\tau, v^0, \bm{k}) \Pith^{nl}(v^0, z^0, \bm{k}) \nonumber \\
		& =      & \Delta^s (-i\tau, \bm{k}) \delta_s (z^0) + \Delta^a (-i\tau, \bm{k}) \delta_a(z^0) \nonumber \\
		&        & {} + \intl_{\I} dv^0 \; D(-i\tau, v^0, \bm{k}) \Pith^{nl} (v^0, z^0, \bm{k}) \nonumber \\
		& =      & \quad \feyn{-1.5mm}{0.175}{DeltaM} \quad + \quad \feyn{-3mm}{0.2}{DpropPi} \nonumber \\
		& \equiv & \quad \feyn{-1.75mm}{0.175}{Delta} \ , \label{ROOE:Delta}
	\end{eqnarray}
	and
	\[ \Delta^T (z^0, -i\tau, \bm{k}) = \Delta (-i\tau, z^0, \bm{k}) = \feynRot{3.5mm}{0.175}{180}{Delta} \;. \]
	The coefficients $\Delta^{s,a}(-i\tau,\bm{k})$ are given by
	\begin{eqnarray}\label{ROOE:DeltaSA}
		\Delta^s(-i\tau,\bm{k}) & = & \frac{\Gth^{\I\I}(-i\tau,0,\bm{k})}{\Gth(0,0,\bm{k})} \;, \nonumber \\
		\Delta^a(-i\tau,\bm{k}) & = & 2 \partial_{\tau} \Gth^{\I\I}(-i\tau,0,\bm{k}) \;.
	\end{eqnarray}
	Eqs.~(\ref{ROOE:GthDelta},\ref{ROOE:Delta},\ref{ROOE:DeltaSA}) constitute the nonperturbative
	generalizations of eqs.~(\ref{GthCDelta},\ref{FreeDelta}).
	
	The nonperturbative generalization of eq.~(\ref{ROOE:GTHFandD0}) is obtained from
	eq.~(\ref{ROOE:DMatsubara}),
	\begin{eqnarray}
		\lefteqn{\Gth^{\I\I}(-i\tau,-i\tau',\bm{k})} \nonumber \\
		& = & D(-i\tau, -i\tau', \bm{k}) + \intl_{\C} dw^0 \intl_{\C} dz^0 \; \Delta(-i\tau, w^0, \bm{k}) \nonumber \\
		&   & {} \times \Gth(w^0, z^0, \bm{k}) \Delta^T (z^0, -i\tau', \bm{k}) \nonumber \\
		& = & D(-i\tau, -i\tau', \bm{k}) \nonumber \\
		&   & {} + \Delta^s(-i\tau, \bm{k}) \Gth(0, 0, \bm{k}) \Delta^s(-i\tau', \bm{k}) \;. \label{ROOE:Dtautau}
	\end{eqnarray}
	Note that only the parts of the connections containing $\Delta^s$ contribute to the double
	integral in the second and third line, whereas the parts involving $\Delta^a$ and
	$\Pith^{\it nl}$ vanish due to a cancellation of the contributions from the two branches of
	the closed real-time path.
	Using the Feynman rules from above, the upper equation can also be written as
	\begin{eqnarray*}
		\feyn{-0.5mm}{0.175}{PropagatorII} & = & \  \feyn{-0.5mm}{0.175}{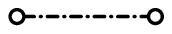} \  + \  \feyn{-1.5mm}{0.4}{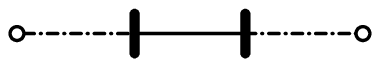}\\
		                                   & = & \  \feyn{-0.5mm}{0.175}{DpropII} \  + \  \feyn{-1.5mm}{0.4}{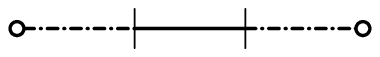}\ .
	\end{eqnarray*}
	
	In Summary, there are two differences compared to the perturbative case:
	(i) the free thermal propagator $\GthF(-i\tau,0,\bm{k})$ enters the free connection, whereas
	the complete thermal propagator $\Gth(-i\tau,0,\bm{k})$ enters the complete connection, and
	(ii) the free connection $\Delta_0(-i\tau,z^0,\bm{k})$ is only supported at the initial time
	$z^0=0_\pm$, whereas the complete connection $\Delta(-i\tau,z^0,\bm{k})$ features an additional
	term containing the non-local part of the complete thermal self-energy.

\end{appendix}


\input{feynRulesCTP.bbl}
\end{document}

%% file: feynRulesCTPInput.tex
\DeclareMathAlphabet{\mathcalbf}{OMS}{cmsy}{b}{n}
\DeclareMathAlphabet{\mathcal}{OMS}{cmsy}{m}{n}

\newcommand{\MPIK}{Max-Planck-Institut f\"ur Kernphysik\\
                   Saupfercheckweg 1, 69117 Heidelberg, Germany\\ \ }

\newcommand{\LRZ}{Leibniz-Rechenzentrum,\\
                  Boltzmannstra\3e 1, 85748 Garching, Germany\\ \ }

\newcommand{\C}{\mathcalbf{C}}

\newcommand{\captionstyle}{\it}

\newcommand{\CpA}{\C\!+\!\alpha}

\newcommand{\CpI}{\C\!+\!\I}

\newcommand{\disc}{\mbox{disc}}

\newcommand{\ds}{\displaystyle}

\newcommand{\EC}{\bm{1_\C}}

\newcommand{\EI}{\bm{1_\I}}

\newcommand{\feyn}[3]{\raisebox{#1}{\includegraphics[width=#2\columnwidth,keepaspectratio]{Pics/Feyn/#3}}}

\newcommand{\feynRot}[4]{\raisebox{#1}{\includegraphics[width=#2\columnwidth,keepaspectratio,angle=#3]{Pics/Feyn/#4}}}

\newcommand{\Gth}{G_{\it th}}

\newcommand{\GthF}{G_{0,{\it th}}}                   

\newcommand{\GthFF}{G_{0,{\it F}}}                   

\newcommand{\GthFR}{G_{0,{\it \rho}}}                

\newcommand{\GthM}{G_{\it m, th}}                    

\newcommand{\GthMF}{G_{\it m, F}}                    

\newcommand{\GthMR}{G_{\it m, \rho}}                 

\newcommand{\I}{\mathcalbf{I}}

\newcommand{\intCI}{\int_{\CpI}\!\!\!\!\!\!\!\!{}}

\newcommand{\intl}{\int\limits}

\newcommand{\KB}{Ka\-da\-noff-Baym\ }

\newcommand{\Pith}{\Pi_{\it th}}

\newcommand{\PithM}{\Pi_{\it m, th}}

\newcommand{\sgn}{\mbox{sign}}

\newcommand{\Tr}{\mbox{Tr}}

\newcommand{\wk}{\omega_{\bm{k}}}

\newcommand{\wks}[1]{\omega_{\bm{k_{#1}}}}

%% file: feynRulesCTP.bbl
\begin{thebibliography}{10}\markboth{\em Bibliography}{\em
  Bibliography}\thispagestyle{empty}
\expandafter\ifx\csname url\endcsname\relax
  \def\url#1{\texttt{#1}}\fi
\expandafter\ifx\csname urlprefix\endcsname\relax\def\urlprefix{URL }\fi
\providecommand{\eprint}[2][]{\url{#2}}

\bibitem{Kofman:1997yn}
Lev Kofman, Andrei~D. Linde, and Alexei~A. Starobinsky, \emph{Towards the
  theory of reheating after inflation}, Phys. Rev. \textbf{D56} (1997) 3258,
\eprint{hep-ph/9704452}.

\bibitem{Kolb:2003dz}
Peter~F. Kolb and Ulrich~W. Heinz, \emph{{Hydrodynamic description of
  ultrarelativistic heavy-ion collisions}}  (2003),
\eprint{nucl-th/0305084}.

\bibitem{Kolb:1990vq}
Edward~W. Kolb and Michael~S. Turner, \emph{The Early universe}  (1990),
  redwood City, USA: Addison-Wesley (1990) 547 p. (Frontiers in physics, 69)

\bibitem{Komatsu:2008hk}
Eiichiro Komatsu et~al. (WMAP), \emph{{Five-Year Wilkinson Microwave Anisotropy
  Probe (WMAP) Observations:Cosmological Interpretation}}, Astrophys. J. Suppl.
  \textbf{180} (2009) 330,
\eprint{0803.0547}.

\bibitem{Berges:2000ur}
J{\"u}rgen Berges and J{\"u}rgen Cox, \emph{Thermalization of quantum fields
  from time-reversal invariant evolution equations}, Phys. Lett. \textbf{B517}
  (2001) 369,
\eprint{hep-ph/0006160}.

\bibitem{Berges:2001fi}
J{\"u}rgen Berges, \emph{Controlled nonperturbative dynamics of quantum fields
  out of equilibrium}, Nucl. Phys. \textbf{A699} (2002) 847,
\eprint{hep-ph/0105311}.

\bibitem{Aarts:2001yn}
Gert Aarts and J{\"u}rgen Berges, \emph{Classical aspects of quantum fields far
  from equilibrium}, Phys. Rev. Lett. \textbf{88} (2002) 041603,
\eprint{hep-ph/0107129}.

\bibitem{Aarts:2003bk}
Gert Aarts and Jose~M. Mart{\'i}nez~Resco, \emph{Transport coefficients from
  the 2{PI} effective action}, Phys. Rev. \textbf{D68} (2003) 085009,
\eprint{hep-ph/0303216}.

\bibitem{Lindner:2005kv}
Manfred Lindner and Markus~Michael M{\"u}ller, \emph{Comparison of {Boltzmann}
  equations with quantum dynamics for scalar fields}, Phys. Rev. \textbf{D73}
  (2006) 125002,
\eprint{hep-ph/0512147}.

\bibitem{Cornwall:1974vz}
John~M. Cornwall, Roman Jackiw, and Eleftherios Tom\-boulis, \emph{Effective
  Action for Composite Operators}, Phys. Rev.
\textbf{D10} (1974) 2428.

\bibitem{Schwinger:1960qe}
Julian~S. Schwinger, \emph{Brownian motion of a quantum oscillator}, J. Math.
  Phys.
\textbf{2} (1961) 407.

\bibitem{Bakshi:1962dv}
Pradip~M. Bakshi and Kalyana~T. Mahanthappa, \emph{{Expectation value formalism
  in quantum field theory. 1}}, J. Math. Phys.
\textbf{4} (1963) 1.

\bibitem{Bakshi:1963bn}
Pradip~M. Bakshi and Kalyana~T. Mahanthappa, \emph{{Expectation value formalism
  in quantum field theory. 2}}, J. Math. Phys.
\textbf{4} (1963) 12.

\bibitem{Keldysh:1964ud}
Leonid~V. Keldysh, \emph{Diagram technique for nonequilibrium processes}, Sov.
  Phys. JETP
\textbf{20} (1965) 1018.

\bibitem{Danielewicz:1982kk}
Pawel Danielewicz, \emph{Quantum Theory of Nonequilibrium Processes {I}},
  Annals Phys.
\textbf{152} (1984) 239.

\bibitem{Berges:2002cz}
J{\"u}rgen Berges and Julien Serreau, \emph{Parametric resonance in quantum
  field theory}, Phys. Rev. Lett. \textbf{91} (2003) 111601,
\eprint{hep-ph/0208070}.

\bibitem{Arrizabalaga:2004iw}
Alejandro Arrizabalaga, Jan Smit, and Anders Tranberg, \emph{Tachyonic
  preheating using 2PI-1/N dynamics and the classical approximation}, JHEP
  \textbf{10} (2004) 017,
\eprint{hep-ph/0409177}.

\bibitem{Aarts:2007qu}
Gert Aarts and Anders Tranberg, \emph{Particle creation and warm inflation},
  Phys. Lett. \textbf{B650} (2007) 65,
\eprint{hep-ph/0701205}.

\bibitem{Aarts:2007ye}
Gert Aarts and Anders Tranberg, \emph{{Thermal effects on slow-roll dynamics}},
  Phys. Rev. \textbf{D77} (2008) 123521,
\eprint{0712.1120}.

\bibitem{Berges:2009bx}
J{\"u}rgen Berges, Jens Pruschke, and Alexander Rothkopf,
  \emph{{Instability-induced fermion production in quantum field theory}}
  (2009),
\eprint{0904.3073}.

\bibitem{Berges:2002wr}
J{\"u}rgen Berges, Szabolcs Borsanyi, and Julien Serreau, \emph{Thermalization
  of fermionic quantum fields}, Nucl. Phys. \textbf{B660} (2003) 51,
\eprint{hep-ph/0212404}.

\bibitem{Berges:2004ce}
J{\"u}rgen Berges, Szabolcs Borsanyi, and Christof Wetterich,
  \emph{Prethermalization}, Phys. Rev. Lett. \textbf{93} (2004) 142002,
\eprint{hep-ph/0403234}.

\bibitem{Juchem:2003bi}
Sascha Juchem, Wolfgang Cassing, and Carsten Greiner, \emph{Quantum dynamics
  and thermalization for out-of-equi\-li\-brium phi**4-theory}, Phys. Rev.
  \textbf{D69} (2004) 025006,
\eprint{hep-ph/0307353}.

\bibitem{Arrizabalaga:2005tf}
Alejandro Arrizabalaga, Jan Smit, and Anders Tranberg, \emph{Equilibration in
  $\varphi^4$ theory in 3+1 dimensions}, Phys. Rev. \textbf{D72} (2005) 025014,
\eprint{hep-ph/0503287}.

\bibitem{Lindner:2007am}
Manfred Lindner and Markus~Michael M{\"u}ller, \emph{{Comparison of {Boltzmann}
  Kinetics with Quantum Dynamics for a Chiral {Yukawa} Model Far From
  Equilibrium}}, Phys. Rev. \textbf{D77} (2008) 025027,
\eprint{arXiv:0710.2917}.

\bibitem{Anisimov:2008dz}
Alexey Anisimov, Wilfried Buchm{\"u}ller, Marco Drewes, and Sebastian
  Mendizabal, \emph{{Nonequilibrium Dynamics of Scalar Fields in a Thermal
  Bath}}  (2008),
\eprint{0812.1934}.

\bibitem{Danielewicz:1982ca}
Pawel Danielewicz, \emph{Quantum theory of nonequilibrium processes. {II}.
  Application to nuclear collisions}, Annals Phys.
\textbf{152} (1984) 305.

\bibitem{Kohler:1995zz}
Sigurd K{\"o}hler, \emph{{Memory and correlation effects in nuclear
  collisions}}, Phys. Rev.
\textbf{C51} (1995) 3232.

\bibitem{Kohler:1996zz}
Sigurd K{\"o}hler, \emph{{Memory and correlation effects in the quantum theory
  of thermalization}}, Phys. Rev.
\textbf{E53} (1996) 3145.

\bibitem{Morawetz:1998em}
Klaus Morawetz and Sigurd K{\"o}hler, \emph{{Formation of correlations and
  energy-conservation at short time scales}}, Eur. Phys. J. \textbf{A4} (1999)
  291,
\eprint{nucl-th/9802082}.

\bibitem{Kohler:2001zv}
Sigurd K{\"o}hler and Klaus Morawetz, \emph{{Correlations in Many-Body Systems
  with Two-time Green's Functions}}, Phys. Rev. \textbf{C64} (2001) 024613,
\eprint{nucl-th/0102059}.

\bibitem{Tranberg:2008ae}
Anders Tranberg, \emph{{Quantum field thermalization in expanding backgrounds}}
   (2008),
\eprint{arXiv:0806.3158}.

\bibitem{Hohenegger:2008zk}
Andreas Hohenegger, Alexander Kartavtsev, and Manfred Lindner, \emph{{Deriving
  Boltzmann Equations from Ka\-da\-noff-Baym Equations in Curved Space-Time}}
  (2008),
\eprint{arXiv:0807.4551}.

\bibitem{Berges:2005hc}
J{\"u}rgen Berges, Szabolcs Borsanyi, Urko Reinosa, and Julien Serreau,
  \emph{Nonperturbative renormalization for 2{PI} effective action techniques},
  Annals Phys. \textbf{320} (2005) 344,
\eprint{hep-ph/0503240}.

\bibitem{Berges:2004hn}
J{\"u}rgen Berges, Szabolcs Borsanyi, Urko Reinosa, and Julien Serreau,
  \emph{Renormalized thermodynamics from the 2{PI} effective action}, Phys.
  Rev. \textbf{D71} (2005) 105004,
\eprint{hep-ph/0409123}.

\bibitem{Blaizot:2003an}
Jean-Paul Blaizot, Edmond Iancu, and Urko Reinosa, \emph{Renormalization of
  phi-derivable approximations in sca\-lar field theories}, Nucl. Phys.
  \textbf{A736} (2004) 149,
\eprint{hep-ph/0312085}.

\bibitem{vanHees:2001pf}
Hendrik van Hees and Joern Knoll, \emph{Renormalization of self-consistent
  approximation schemes. {II}: Applications to the sunset diagram}, Phys. Rev.
  \textbf{D65} (2002) 105005,
\eprint{hep-ph/0111193}.

\bibitem{vanHees:2001ik}
Hendrik van Hees and Joern Knoll, \emph{Renormalization in self-consistent
  approximations schemes at finite temperature. {I}: Theory}, Phys. Rev.
  \textbf{D65} (2002) 025010,
\eprint{hep-ph/0107200}.

\bibitem{Borsanyi:2008ar}
Szabolcs Borsanyi and Urko Reinosa, \emph{{Renormalised non\-equilibrium
  quantum field theory: scalar fields}}  (2008),
\eprint{0809.0496}.

\bibitem{Niemi:1983ea}
Antti~J. Niemi and Gordon~W. Semenoff, \emph{{Thermodynamic Calculations in
  Relativistic Finite Temperature Quantum Field Theories}}, Nucl. Phys.
\textbf{B230} (1984) 181.

\bibitem{Niemi:1983nf}
Antti~J. Niemi and Gordon~W. Semenoff, \emph{{Finite Temperature Quantum Field
  Theory in Minkowski Space}}, Ann. Phys.
\textbf{152} (1984) 105.

\bibitem{Landsman:1986uw}
Nicolaas~P. Landsman and Christianus~G. van Weert, \emph{{Real and Imaginary
  Time Field Theory at Finite Temperature and Density}}, Phys. Rept.
\textbf{145} (1987) 141.

\bibitem{Chou:1984es}
Kuangchao Chou, Zhaobin Su, Bailin Hao, and Lu~Yu, \emph{{Equilibrium and
  Nonequilibrium Formalisms Made Unified}}, Phys. Rept.
\textbf{118} (1985) 1.

\bibitem{Calzetta:1986cq}
Esteban Calzetta and Bei-Lok Hu, \emph{Nonequilibrium quantum fields: closed
  time path effective action, {Wigner} function and {Boltzmann} equation},
  Phys. Rev.
\textbf{D37} (1988) 2878.

\bibitem{LeBellac:1996at}
Michel Le~Bellac and H.~Mabilat, \emph{{Real time Feynman rules at finite
  temperature}}, Phys. Lett.
\textbf{B381} (1996) 262.

\bibitem{Gelis:1994dp}
Francois Gelis, \emph{The Effect of the vertical part of the path on the real
  time {Feynman} rules in finite temperature field theory}, Z. Phys.
  \textbf{C70} (1996) 321,
\eprint{hep-ph/9412347}.

\bibitem{Gelis:1999nx}
Francois Gelis, \emph{{A new approach for the vertical part of the contour in
  thermal field theories}}, Phys. Lett. \textbf{B455} (1999) 205,
\eprint{hep-ph/9901263}.

\bibitem{Berges:2004yj}
J{\"u}rgen Berges, \emph{Introduction to nonequilibrium quantum field theory},
  AIP Conf. Proc. \textbf{739} (2005) 3,
\eprint{hep-ph/0409233}.

\bibitem{Garny:PhD}
Mathias Garny, \emph{Particle Physics and Dark Energy: Beyond Classical
  Dynamics}, Ph.D. thesis, Munich, Tech. U. (2008)

\bibitem{Berges:2004pu}
J{\"u}rgen Berges, \emph{n-{PI} effective action techniques for gauge
  theories}, Phys. Rev. \textbf{D70} (2004) 105010,
\eprint{hep-ph/0401172}.

\end{thebibliography}
